\documentclass[a4paper,english,a4paper,oneside]{amsart}
\usepackage[T1]{fontenc}
\usepackage[latin1]{inputenc}
\usepackage{graphicx}
\usepackage{amssymb}

\makeatletter
 \theoremstyle{plain}
\newtheorem{thm}{Theorem}[section]
  \theoremstyle{plain}
  \newtheorem{lem}[thm]{Lemma}
  \theoremstyle{remark}
  \newtheorem{rem}[thm]{Remark}

\usepackage{babel}
\makeatother
\begin{document}

\title{On the Stopping Time of a Bouncing Ball}

\author{Anna Maria Cherubini, Giorgio Metafune, Francesco Paparella\\%
Dip. di Matematica E. De Giorgi - Università di Lecce - Lecce, Italy.}

\email{anna.cherubini@unile.it, giorgio.metafune@unile.it, francesco.paparella@unile.it}

\begin{abstract}
We study a simple model of a bouncing ball that takes explicitely
into account the elastic deformability of the body and the energy
dissipation due to internal friction. We show that this model is not
subject to the problem of inelastic collapse, that is, it does not
allow an infinite number of impacts in a finite time. We compute asymptotic
expressions for the time of flight and for the impact velocity. We
also prove that contacts with zero velocity of the lower end of the
ball are possible, but non-generic. Finally, we compare our findings
with other models and laboratory experiments. 
\end{abstract}
\maketitle

\section{Introduction}

In this paper we study how a ball bouncing against an horizontal,
rigid plane comes to rest. We are motivated by the fact that bouncing
objects are the basic building blocks of granular fluids. Our view
is that an acceptable mathematical description of granular fluids
is impossible without taking into account internal vibrations of the
bouncing objects. To explore the viability of this program we formulate
a simple, one-dimensional model that explicitly accounts for the deformation
of the ball, and we prove that, while encompassing the desirable properties
of current models, which neglect internal vibrations, it does not
incur their pathologies.

The simplest and most widely used model of a bouncing ball (or grains
of a granular fluid) assumes that the ball is a rigid body, and that
an impact with the floor is an instantaneous event, which reverses
the vertical component of the speed of the ball. In order to model
energy dissipation caused by an impact, it is customary to introduce
a positive coefficient of restitution $r<1$, so that the vertical
speed $u_{a}$ immediately after an impact is related to the vertical
speed $u_{b}$ immediately before the impact by the simple relationship
\begin{equation}
u_{a}=-ru_{b}.\label{eq:standard-model}\end{equation}
 This model performs well in cases where the ball does not experience
too many impacts in the unit of time. It has been used, for example,
as an ingredient in the description of sport balls \cite{Cross},
and to study the dynamics of a bead on a vibrating plate (for its
deep mathematical facetes the latter has become a classical problem,
see \cite{GuckHol} sec. 2.4). The assumptions behind this model (namely:
neglect of deformability, instantaneous impacts and energy losses
described by a restitution coefficient) are the basic building blocks
of current theories of granular gases \cite{Brilliantovbook}.

The most apparent drawback in this approach is that it cannot limit
\emph{a priori} the number of impacts in the unit of time. In fact,
granular systems described by (\ref{eq:standard-model}), for a large
class of parameter choices and initial conditions, are subject to
the phenomenon of \emph{inelastic collapse}, where clusters of particles
are subject to an infinite number of collisions in a finite time \cite{McNamara-Young}.
The simplest example of inelastic collapse is given by a single ball,
subject to a constant gravity force, bouncing repeatedly off the floor.
Neglecting the interaction with the air, the vertical speed of the
ball immediately after the $(n+1)-$th impact is linked to that at
the $n-$th impact by\begin{equation}
u_{n+1}=ru_{n}.\label{eq:geometric-map}\end{equation}
 The duration of the $n-$th flight is $\tau_{n}=2u_{n}/g$, where
$g$ is the acceleration of gravity. The sum of the times of flight
is easily computed, and gives a geometric series that converges to\[
t_{\infty}=\frac{2u_{0}}{g(1-r)}.\]
 For times larger than $t_{\infty}$ the model is meaningless.

This pathology is not caused by the one-dimensional nature of the
model sketched here. Even taking into account rotational degrees of
freedom and exchanges of angular momentum at the impacts due to surface
friction, the inelastic collapse is still a common outcome \cite{SchorghoferZhou}.

To avoid the problem it is customary to assume that the restitution
coefficient is an increasing function of the impact velocity, and
that $r(u)\rightarrow1$ if $u\rightarrow0$. A popular choice is\begin{equation}
r(u)=\left(1-\left(\frac{u}{U}\right)^{1/5}\right)+O\left(\left(\frac{u}{U}\right)^{2/5}\right).\label{eq:onefifth-rest-coeff}\end{equation}
 where $U$ is a constant that depends on the material and geometry
of the ball. This expression has some theoretical support \cite{Ramirez},
and it appears to fit the data for non-repeated impacts \cite{Bridges84}.
In the absence of gravity, inelastic collapse is ruled out rigorously
for systems of three balls \cite{Goldman}, and there is numerical
evidence that the same result holds for an arbitrary number of particles
(see \cite{Brilliantovbook} chap. 26 and references therein). However,
using (\ref{eq:onefifth-rest-coeff}) in (\ref{eq:geometric-map}),
yields again a converging sum of the times of flight for the bouncing
ball problem. This is proved rigorously in the appendix (Remark \ref{rem:x-xbeta-map}).

The common wisdom is that $t_{\infty}$ is the time when the bouncing
ball comes to rest and maintains permanent contact with the floor
(see, for example \cite{Brilliantovbook} chap. 3). This point of
view faces serious difficulties if one desires to model, for example,
the settling-down of many beads poured into a box: it is not clear
how a bead at rest on the floor should behave when hit by a moving
one, because there is no obvious way to extend (\ref{eq:standard-model})
to situations where three or more bodies are in contact simultaneously.

More generally, the assumption (\ref{eq:standard-model}) rests on
hypoteses that become invalid as the frequency of the impacts diverges:
collisions with the floor are not truly instantaneous, and treating
the ball as indeformable is highly questionable when the frequency
of impacts is close to the resonant frequencies of the bouncing ball
(e.g. \cite{Duran99} sec. 2.2).

Of course, a model avoiding inelastic collapse must at the same time
account for the obvious observation that bouncing balls do come to
rest after a finite time. A first step in this direction is the time-of-contact
model \cite{Luding-McNamara}, which prescribes $r<1$ if the time-of-flight
of the ball is greater than a constant $T_{c}$, and $r=1$ otherwise:
conservative impacts are interpreted as internal vibrations of the
ball, which (from a macroscopic point of view) maintains contact with
the floor. A variation on this theme leads to the notion of a stochastic
restitution coefficient \cite{Zippelius}.

In this paper we study a one-dimensional model of a bouncing ball
simple enough to allow for rigorous mathematical analysis, but including
all the elements that we believe are important for further developments
in the description of granular fluids. We explicitly take into account
the deformability of the body, and at the same time we give up the
notion of restitution coefficient, at least as a primitive concept.
We prove that our model is free from pathologies analogous to the
inelastic collapse, and that it is well defined at all times. We still
assume impacts to be instantaneous, but only from a microscopic point
of view: a persistent contact with the floor is seen as a rapid sequence
of instantaneous impacts. This property makes the dynamics only piecewise-smooth.
There is a growing literature on piecewise smooth dynamical systems,
with engineering and plasma physics applications (for a review see
\cite{Budd-book}). However, the focus is generally on periodically
forced systems, rather than on the pathways chosen by an unforced
system to reach the asymptotic equilibrium.

The rest of the paper is organized as follows: in section \ref{sec:model-description}
we describe the model and we state the main results: in particular
the absence of inelastic collapse and how the system approaches the
rest state; the theorems are proved in sections \ref{sec:Sticky-Events}
through \ref{sec:ZeroMeasure}; in section \ref{sec:Numerical-Results}
we show that the notion of restitution coefficient is naturally recovered,
as a consequence of the dynamics, when the time-of-flight is large
with respect to the characteristic damping time of the internal vibrations;
numerical simulations are compared with the laboratory experiments
of \cite{Falcon98}; finally, section \ref{sec:Discussion-and-Conclusions}
contains a summary of the results and some forward-looking remarks.

\section{\label{sec:model-description}Equations of Motion}

An basic model of a deformable bouncing ball is shown in figure%
\begin{figure}
\begin{centering}\includegraphics[width=0.9\columnwidth]{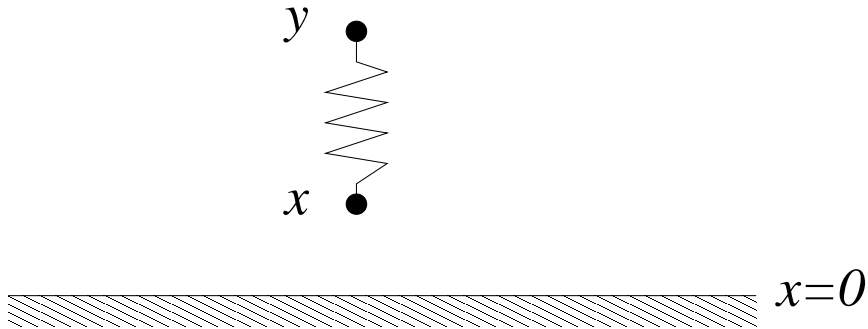} \par\end{centering}

\caption{\label{fig:A-bouncing-ball}A bouncing ball of mass $2m$ and diameter
$L$ is idealized as two points of mass $m$ connected by a massless,
dissipative spring.}
\end{figure}
 (\ref{fig:A-bouncing-ball}): two point masses are connected by a
massless dissipative spring. This idealized ball, when it is not in
contact with the floor, is ruled by the following equations of motion
\begin{equation}
\left\{ \begin{array}{c}
m\ddot{x}=-gm+k(y-x-L)+\nu(\dot{y}-\dot{x})\\
m\ddot{y}=-gm-k(y-x-L)-\nu(\dot{y}-\dot{x})\end{array}\right.\label{eq:dimensional-eq-motion}\end{equation}
 where $m$ is the mass of the material points at $x(t)$ and $y(t)$,
$L$ is the length at rest of the spring, $k$ is its elastic constant,
$\nu$ is a damping coefficient, and $-g$ is the acceleration of
gravity. All constants are positive.

We indicate by $t_{n}$ the time when an impact occurs, i.e. $x(t_{n})=0$.
We define the \emph{time of flight} \begin{equation}
\tau_{n}=t_{n+1}-t_{n}\label{eq:time-of-flight}\end{equation}
 and we assume $t_{0}=0$. Impacts are modeled as an instantaneous
elastic interaction obeying to the rule\begin{equation}
\dot{x}(t_{n}^{+})=-\dot{x}(t_{n}^{-}).\label{eq:collision-rule-x1}\end{equation}
 where the notation $f(a^{\pm})$ means $\lim_{t\rightarrow a^{\pm}}f(t)$.
The boundary does not exert any force directly on the mass in $y$,
which is affected by boundary hits only through the resulting deformation
of the spring. In other words, at the impact times $t_{n}$ the upper
point mass obeys the rule\begin{equation}
\dot{y}(t_{n}^{+})=\dot{y}(t_{n}^{-}).\label{eq:collision-rule-y1}\end{equation}
 The special case $\dot{x}_{n}(t_{n})=0$ may lead to a continuous
contact of the lower point mass with the floor. This is discussed
in detail in the next section.

We re-write the equations (\ref{eq:dimensional-eq-motion}) using
$L$ as the scale of lengths, and $\sqrt{m/(2k)}$ as the scale of
times. The resulting dimensionless equations are\begin{equation}
\left\{ \begin{array}{c}
\ddot{x}+\mu\dot{x}+\frac{1}{2}x=-\gamma-\frac{1}{2}+\frac{1}{2}y+\mu\dot{y}\\
\ddot{y}+\mu\dot{y}+\frac{1}{2}y=-\gamma+\frac{1}{2}+\frac{1}{2}x+\mu\dot{x}\end{array}\right.\label{eq:dimensionless-eq-motion}\end{equation}
 where $\gamma=gm/(2Lk)$ and $\mu=\nu/\sqrt{2km}$.

The positions $x(t)$ and $y(t)$ are defined for $t\in[0,t_{\infty})$,
where $t_{\infty}$ is the time of inelastic collapse, should it occur,
or $\infty$ otherwise. For $t\in\cup_{n}(t_{n},t_{n+1})$, the equations
(\ref{eq:dimensionless-eq-motion}) guarantee the existence of the
derivatives $x^{(k)}$ and $y^{(k)}$ for all $k$. For brevity, it
is convenient to define $Y_{n}=Y(t_{n})$, $\dot{X}_{n}=\lim_{t\rightarrow t_{n}^{+}}\dot{x}(t)$,
$\dot{Y}_{n}=\lim_{t\rightarrow t_{n}^{+}}\dot{y}(t)$, and similarly
for all higher derivatives. Of course it is $X_{n}=x(t_{n})=0$ for
all $n$. With this notation, the collision rule (\ref{eq:collision-rule-x1})
becomes \begin{equation}
\dot{X}_{n}=-\lim_{t\rightarrow t_{n}^{-}}\dot{x}(t).\label{eq:collision-rule-map}\end{equation}

Next we define the variables\begin{equation}
\left\{ \begin{array}{c}
\psi=(y+x-1)/2\\
\xi=(y-x-1)/2\end{array}\right.\label{eq:xi-psi}\end{equation}
 The collision condition $x(t_{n})=0$ in the new variables reads
\begin{equation}
\psi(t_{n})=\xi(t_{n}).\label{eq:collision-condition-psi-xi}\end{equation}
 The collision rule (\ref{eq:collision-rule-x1}) becomes\begin{equation}
\left\{ \begin{array}{c}
\dot{\psi}(t_{n}^{+})=\dot{\xi}(t_{n}^{-})\\
\dot{\xi}(t_{n}^{+})=\dot{\psi}(t_{n}^{-})\end{array}\right..\label{eq:collision-rule}\end{equation}
 In terms of these variables the equations of motion read\begin{equation}
\left\{ \begin{array}{c}
\ddot{\psi}=-\gamma\\
\ddot{\xi}=-\xi-2\mu\dot{\xi}\end{array}\right..\label{eq:eq-motion-psi-xi}\end{equation}

The total mechanical energy of the idealized ball is \begin{equation}
E=\frac{\dot{\xi}^{2}}{2}+\frac{\xi^{2}}{2}+\frac{\dot{\psi}^{2}}{2}+\gamma\psi.\label{eq:tot-mech-energy}\end{equation}
 which is dissipated at the rate \begin{equation}
\frac{dE}{dt}=-2\mu\dot{\xi}^{2}.\label{eq:energy-dissipation}\end{equation}
 The collision rule (\ref{eq:collision-rule}) implies that $E$ is
a continuous function of time, even at the impact times. The energy
is then a non-increasing function of time.

The mechanical system has only one static equilibrium, which is \begin{equation}
x=\dot{x}=\dot{y}=0,\quad y=1-2\gamma.\label{eq:equilibrium}\end{equation}
 This state has energy\begin{equation}
E_{m}=-\gamma^{2}/2\label{eq:Emin}\end{equation}
 which is the minimal energy of the system.

To insure that the upper point mass at equilibrium is above the floor,
we must require that \begin{equation}
\gamma<\frac{1}{2}.\label{eq:gamma<0.5}\end{equation}

The validity of the model may be questioned if the spring's length
shrinks to zero, that is if, for some $\bar{t}$, we find $x(\bar{t})=y(\bar{t})$,
or, equivalently, $\xi(\bar{t})=-1/2$. To avoid this event we may
choose initial conditions having mechanical energy less than the minimum
energy stored in a ball with zero length spring, that is \begin{equation}
E<\frac{1-4\gamma}{8}.\label{eq:max-ini-energy}\end{equation}
 Imposing an upper bound to the mechanical energy translates the fact
that a real ball thrown on a rigid floor with excessive energy would
break or undergo plastic deformations, thus changing the physics of
the problem. However, from a mathematical standpoint, we may wish
to study the case where the constraint represented by the floor applies
only to the lower point mass. That is, we require $x(t)\ge0$ for
any $t$, but we allow $y(t)$ to be negative. In this paper we need
not to enforce the restriction (\ref{eq:gamma<0.5}) and (\ref{eq:max-ini-energy}),
and the only constraint is $x(t)\ge0$.

The main result is the following theorem, which rules out the possibility
of inelastic collapse in our model.

\begin{thm}
\label{thm:main_thm}Starting from any initial condition, there are
two possible outcomes as $t\rightarrow\infty$: either the lower point
mass remains in contact with the floor, while the upper one undergoes
damped harmonic oscillations; or the lower point mass experiences
an unlimited number of instantaneous impacts with the floor, in which
case the times of flight (\ref{eq:time-of-flight}) follow the asymptotic
relation \[
\tau_{n}\sim\frac{3}{\mu}\frac{1}{n}.\]

\end{thm}
Although in both cases our model is well defined for any positive
time, the first outcome is non-generic. In fact, we shall prove that

\begin{thm}
\label{thm:contact-non-generic}The contacts of the lower point mass
with the floor are always instantaneous, that is $x(t_{n})=0$ implies
$\lim_{t\rightarrow t_{n}^{-}}\dot{x}(t)\ne0$, except for the solutions
generated by a set of initial conditions having zero Lebesgue measure.
Furthermore, this set is nowhere dense in the set of all possible
initial conditions.
\end{thm}
A by-product of the main theorem is the following

\begin{thm}
\label{thm:static-equilibrium}Starting from any initial condition,
the mechanical system described in this section tends to the state
of static equilibrium (\ref{eq:equilibrium}) as $t\rightarrow\infty$. 
\end{thm}

\section{\label{sec:Sticky-Events}Anomalous Contacts}

The design goal of the mechanical system described in section (\ref{sec:model-description})
was modelling a prolonged contact between a ball and the floor with
a sequence of instantaneous impacts, where the lower point mass reaches
the floor with a non-zero velocity. However, there may be some anomalous
contacts where this is not the case: their properties need to be fully
understood before proving the theorems stated in the previous section,
even if we will show that they arise only from a zero measure set
of initial conditions.

The first class of such contacts, that are often named \emph{grazings}
in the literature, are instantaneous contacts where there is no exchange
of momentum between the lower point mass and the floor. A grazing
event occurs when \begin{equation}
x(t_{s})=\dot{x}(t_{s})=0,\qquad\ddot{x}(t_{s})>0\label{eq:swishes}\end{equation}
 at some time $t_{s}$. Applying the impact rule (\ref{eq:collision-rule-x1})
we realize that the presence of the floor is irrelevant during a grazing,
because in an interval around $t_{s}$ the dynamics would be the same
with of without the floor. We note that a trajectory can not have
$x(t_{s})=\dot{x}(t_{s})=0,$ $\ddot{x}(t_{s})<0$, because that would
violate the constraint $x(t)\ge0$ for times close to $t_{s}$.

On the other hand, it is possible to have trajectories touching the
floor with zero velocity and acceleration, and this leads to the second
class of anomalous contacts, where the lower point mass maintains
contact with the floor for a non-zero interval of time. We will call
this one a \emph{sticky event.} The duration of a sticky event can
either be infinite, leading to the first of the two outcomes mentioned
in Theorem (\ref{thm:main_thm}), or it may be finite, then the mechanical
system resumes its ordinary dynamics with instantaneous impacts.

A \emph{sticky event} begins at a certain time $t_{s}$ if \begin{equation}
x(t_{s})=\dot{x}(t_{s})=\ddot{x}(t_{s})=0.\label{eq:sticky-condition}\end{equation}
 In such a case $\dddot{x}(t_{s})<0$. In fact, $\dddot{x}(t_{s})>0$
is inconsistent with the constraint $x(t)\ge0$ for $t$ approaching
$t_{s}$ from below. Moreover, by differentiating twice the first
of (\ref{eq:dimensionless-eq-motion}) and using the second to eliminate
$y$ and its derivatives, one finds that $\dddot{x}(t_{s})=0$ implies
$x^{(4)}(t_{s})=-\gamma$, which also is inconsistent with the constraint.

In this situation the collision rule (\ref{eq:collision-rule-x1})
does not change the state of the system but, to avoid breaking the
constraint $x(t)\ge0$, it is clear that a new element must come into
play, namely a non-instantaneous force exerted by the floor onto the
lower point mass.

Using the equations of motion (\ref{eq:dimensionless-eq-motion})
and their first derivatives, we find that conditions (\ref{eq:sticky-condition})
are equivalent to \begin{equation}
X_{s}=\dot{X}_{s}=0,\quad Y_{s}=1+2\gamma-2\mu\dot{Y}_{s}\label{eq:sticky-cond-Y}\end{equation}
 where, for brevity, we define $X_{s}=x(t_{s}),$ $Y_{s}=y(t_{s}),$
etc. Moreover, $\dddot{x}(t_{s})<0$ becomes $\dot{Y}_{s}<4\gamma\mu$.

The sum of the (non-dimensional) forces exerted by the spring and
by gravity on the lower point mass is \begin{equation}
F=\frac{1}{2}y+\mu\dot{y}-\gamma-\frac{1}{2}.\label{eq:force-on-the-floor}\end{equation}
 We observe that $F(t_{s})=0$ and $\dot{F}(t_{s})=\dddot{x}(t_{s})<0$.
During a sticky event $F$ is non-positive and it is balanced by the
reaction exerted by the floor. The lower point mass maintains contact
with the floor until the time $t_{d}>t_{s}$ where the following conditions
are satisfied

\begin{equation}
\left\{ \begin{array}{ccc}
F(t_{d}) & = & 0\\
\dot{F}(t_{d}) & > & 0\end{array}\right..\label{eq:cond_di_distacco}\end{equation}
 The second condition has to be a strict inequality because $F(t)=\dot{F}(t)=0$
easily implies $\ddot{F}(t)=-\gamma$, using (\ref{eq:y-eq-sticky})
below, and this means that the force remains non-positive around such
a $t$. In the interval $[t_{s},t_{d}]$ the motion of the upper point
mass obeys the equation\begin{equation}
\ddot{y}+\mu\dot{y}+\frac{1}{2}y=-\gamma+\frac{1}{2}.\label{eq:y-eq-sticky}\end{equation}

The following lemma is pivotal in order to understand the properties
of anomalous contacts.

\begin{lem}
\label{lem:Min-sticky-energy}Sticky and grazing events are impossible
if the energy of the mechanical system is less than \begin{equation}
E_{min}=\frac{3-2\mu^{2}}{2(1+2\mu^{2})}\gamma^{2}.\label{eq:min-sticky-energy}\end{equation}

\end{lem}
\begin{proof}
Using (\ref{eq:tot-mech-energy}), (\ref{eq:swishes}) or (\ref{eq:sticky-cond-Y}),
and defining $z=Y_{s}+2\gamma-1$, we find that the mechanical energy
at the onset of a sticky event or a grazing is

\begin{equation}
E(t_{s})=\frac{\dot{Y}_{s}^{2}}{4}+\frac{z^{2}}{8}-\frac{\gamma^{2}}{2}.\label{eq:sticky-energy}\end{equation}
 We further have the inequality \begin{equation}
z\ge4\gamma-2\mu\dot{Y}_{s}\label{eq:z-constraint}\end{equation}
 where the equality holds for sticky events and the strict inequality
for grazings. The energy (\ref{eq:sticky-energy}) subject to the
constraint (\ref{eq:z-constraint}) reaches its minimum for\begin{equation}
\dot{Y}_{s}=\frac{4\gamma\mu}{1+2\mu^{2}}\label{eq:ydot-min-Esticky}\end{equation}
 and\[
z=\frac{4\gamma}{1+2\mu^{2}}\]
 where the energy assumes the value (\ref{eq:min-sticky-energy}).
For lower energies the conditions (\ref{eq:swishes}) and (\ref{eq:sticky-cond-Y})
can not be satisfied, and this forbids the occurrence of an anomalous
contact. 
\end{proof}
We observe that $E_{min}>E_{m}$, the absolute minimum energy of the
ball as defined in (\ref{eq:Emin}), and the coincide only in the
limit $\mu\rightarrow\infty$. For all finite values of $\mu$ there
is a finite interval of energies corresponding to states that can
not give rise to anomalous contacts.

The next lemma bounds from below the duration of a sticky event.

\begin{lem}
\label{lem:td-ts-bound}The duration $t_{d}-t_{s}$ of a sticky event
can not be made arbitrarily short. 
\end{lem}
\begin{proof}
Observe that $F$, defined by (\ref{eq:force-on-the-floor}) in terms
of the solution of (\ref{eq:y-eq-sticky}), is a $C^{\infty}$ function
in the interval $[t_{s},t_{d}]$. Let us say that, with a suitable
choice of $\dot{Y}_{s}$, the duration $\zeta_{s}=t_{d}-t_{s}$ may
be made arbitrarily small. Then $\zeta_{s}$ satisfies the equation
\[
0=F(t_{d})=F(t_{s})+\dot{F}(t_{s})\zeta_{s}+\frac{1}{2}\ddot{F}(t_{s})\zeta_{s}^{2}+O(\zeta_{s}^{3}).\]
 Since $F(t_{s})=0$ and $\zeta_{s}>0$, using (\ref{eq:y-eq-sticky})
and its derivatives together with (\ref{eq:sticky-cond-Y}) in order
to express $\dot{F}(t_{s})$ and $\ddot{F}(t_{s})$ in terms of $\dot{Y}_{s}$,
we get \[
2(\dot{Y}_{s}-4\gamma\mu)=(2\gamma+\dot{Y}_{s}\mu-4\gamma\mu^{2})\zeta_{s}+O(\zeta_{s}^{2}).\]
 Setting $\alpha_{s}=4\gamma\mu-\dot{Y}_{s}>0$ we write the above
expression in the form \begin{equation}
2\alpha_{s}+(2\gamma-\mu\alpha_{s})\zeta_{s}=O(\zeta_{s}^{2}).\label{eq:contradictory-td-ts}\end{equation}
 Then $\alpha_{s}\to0$ if $\zeta_{s}\to0$ and both terms in the
left hand side of (\ref{eq:contradictory-td-ts}) are positive, which
leads to the contradiction $\zeta_{s}=O(\zeta_{s})^{2}$. 
\end{proof}
Let us stress that an upper bound on the duration of a sticky event
does not exist. A heuristic argument to support this statement is
the following: if we take an initial condition where only $y$ and
$\dot{y}$ are just slightly removed from the static equilibrium (\ref{thm:static-equilibrium}),
we expect the force $F$ to be always negative, which means that the
system evolves forever according to (\ref{eq:y-eq-sticky}). On the
other hand, if we evolved backward in time this initial condition,
an increasing amount of energy would accumulate in the spring, until
the ball had to bounce above the floor. This means that there is a
family of states, not in contact with the floor, that give rise to
a sticky events that last forever. With tedious, but easy calculations,
it is possible to verify that the state $(x,\dot{x},y,\dot{y})=(0,0,1+2\gamma-2\mu\dot{y},\dot{y})$
with $\dot{y}=4\mu\gamma-\epsilon$, is the onset of such an infinitely
long sticky event, for sufficiently small positive $\epsilon$.

We can now prove the following

\begin{thm}
\label{thm:Finite-number-sticky-ev}Any initial condition generates
a solution which contains at most a finite number of sticky events. 
\end{thm}
\begin{proof}
Assume, by contradiction, that infinitely many sticky events occur
at time intervals $[s_{n},t_{n}]$. Clearly $t_{n}<s_{n+1}$ and Lemma
(\ref{lem:td-ts-bound}) guarantees that there is a positive $\bar{t}$
smaller than any $t_{n}-s_{n}$.

The energy dissipated during the $n^{th}$-sticky event is \[
-\mu\int_{s_{n}}^{t_{n}}\dot{y}^{2}(s)\, ds.\]
 If $|\dot{Y}_{s_{n}}|\ge\delta>0$ for infinitely many $n$, using
the boundedness of $\ddot{y}$, one sees that there exists $0<\hat{t}\le\bar{t}$,
independent of $n$, such that $|\dot{y}(s)|\ge\delta/2$ in the time
interval $[s_{n},s_{n}+\hat{t}]$. This implies that the system dissipates
infinitely many times a fixed amount of energy and contradicts Lemma
(\ref{lem:Min-sticky-energy}). Then $\dot{Y}_{s_{n}}\to0$.

We expand $\dot{y}^{2}$ near $s_{n}$ as \[
\dot{y}^{2}(s_{n}+t)=\dot{Y}_{s_{n}}^{2}t+\dot{Y}_{s_{n}}\ddot{Y}_{s_{n}}t^{2}+\frac{1}{3}\left(\ddot{Y}_{s_{n}}^{2}+\dot{Y}_{s_{n}}\dddot{Y}_{s_{n}}\right)t^{3}+O\left(t^{4}\right)\]
 and we observe that (\ref{eq:y-eq-sticky}) and (\ref{eq:sticky-cond-Y})
yield $\ddot{Y}_{s_{n}}=-2\gamma$. Setting $\epsilon_{n}=\dot{Y}_{s_{n}}$
it follows that there exists $0<\hat{t}\le\bar{t}$ and independent
of $n$ such that \[
\dot{y}^{2}(s_{n}+t)\ge at^{3}-C\epsilon_{n}t\]
 in $[s_{n},s_{n}+\hat{t}]$, for suitable $a,C>0$ (any $a<16/3\gamma^{2}$
works). Then the system dissipates at least $b\hat{t}^{4}-D\epsilon_{n}\hat{t}^{2}$,
$b=a/4,$ $D=C/2$, during the $n^{th}$ sticky event. Since $\epsilon_{n}\to0$,
if $n$ is big enough this quantity is bounded from below and, as
before, we are in contradiction with Lemma (\ref{lem:Min-sticky-energy}). 
\end{proof}
\begin{thm}
\label{thm:Finite-number-swishes}Any initial condition generates
a solution which contains at most a finite number of grazings. 
\end{thm}
The proof of this theorem is trivial, once the truth of (\ref{thm:main_thm})
and (\ref{thm:static-equilibrium}) has been proven. For coherence
with the topic of this section, we give it here, rather than at the
end of section \ref{sec:Asymptotic-State}. Of course, this theorem
is not used in the following two sections.

\begin{proof}
The absence of inelastic collapse (section \ref{sec:Proof-main-th})
forbids an infinite sequence of grazings taking place in a finite
time. By Theorem (\ref{thm:static-equilibrium}) the asymptotic state
is the static equilibrium (\ref{eq:equilibrium}) to which corresponds
an energy smaller than the minimal energy (\ref{eq:min-sticky-energy})
needed for a graze. Then an infinite sequence of grazings in an infinite
time is also forbidden. 
\end{proof}

\section{\label{sec:Proof-main-th}Absence of Inelastic Collapse}

\subsection{Preliminary Considerations}

In this section we prove that the idealized ball described in section
\ref{sec:model-description} does not experience infinite contacts
with the floor in a finite time. The proof focuses on true impacts,
that is contacts such that $\dot{X}_{n}\neq0$, so we must briefly
discuss the case $\dot{X}_{n}=0$.

As we have already seen, this leads to anomalous contacts, either
grazings or sticky events. In the latter case, if the duration of
the event is infinite, the problem of inelastic collapse is avoided,
because equation (\ref{eq:y-eq-sticky}) holds forever. Otherwise,
the discussion of this section may be thought as taking place after
the last sticky event, thanks to Theorem (\ref{thm:Finite-number-sticky-ev})
which allows only a finite number of such events.

Grazings, too, may not happen after an arbitrarily high number of
contacts, but at this stage there isn't a simple and quick way to
prove this statement. Instead, for each of the cases considered below
we will take care to show that $\dot{X}_{n}>0$ for $n$ large enough.

\subsection{Asymptotic relationships at collapse}

In order to show that there is no inelastic collapse, we assume, by
contradiction, that $t_{\infty}$ is a finite positive number. This
assumption leads to explicit approximate expressions for the function
$y$ and its derivative for $t$ close to $t_{\infty}$. In turn,
these expressions allow to deduce asymptotic laws for $\dot{X}_{n}$,
$\ddot{X}_{n}$, $\dddot{X}_{n}$, and $\tau_{n}$ which lead to contradictions.

From (\ref{eq:energy-dissipation}) we have that the mechanical energy
of the system is a non-increasing function of time, and, taking into
account that it can never decrease below the energy (\ref{eq:Emin})
of the static equilibrium (\ref{eq:equilibrium}) we conclude that
the mechanical energy has a finite limit. The existence of such a
limit implies that the dynamical variables $\psi,$ $\dot{\psi},$
$\xi$, $\dot{\xi}$, appearing in the expression (\ref{eq:tot-mech-energy})
for $E$, are bounded, and so are their linear combinations $x,$
$y,$ $\dot{x},$ $\dot{y}$.

The equations of motion (\ref{eq:dimensionless-eq-motion}) guarantee
that $\ddot{x}$ and $\ddot{y}$ are also bounded. In general, iterated
differentiations of (\ref{eq:dimensionless-eq-motion}) show there
exist positive constants $c_{k}$ such that\begin{equation}
\left|x^{(k)}(t)\right|,\left|y^{(k)}(t)\right|<c_{k}\label{eq:x-y-boundedness}\end{equation}
 for every $t\in\cup_{n}(t_{n},t_{n+1})$.

By construction, the positions $x$ and $y$ of the point masses are
continuous functions of time. The collision rule (\ref{eq:collision-rule-x1})
makes $\dot{x}$ discontinuous at the impact times $t_{n}$, while
the rule (\ref{eq:collision-rule-y1}) states that $\dot{y}$ is continuous.
Then $x\in C^{0}[0,t_{\infty})$ and $y\in C^{1}[0,t_{\infty})$.
Furthermore, the limits $Y_{\infty}=\lim_{t\rightarrow t_{\infty}}y(t)$
and $\dot{Y}_{\infty}=\lim_{t\rightarrow t_{\infty}}\dot{y}(t)$ exist
and are finite, because $y\in C^{1}[0,t_{\infty})$ and $\ddot{y}$
is bounded.

Now we prove that position and velocity of the lower point mass go
to zero as $t\rightarrow t_{\infty}$. Integrating the first equation
in (\ref{eq:eq-motion-psi-xi}) and using (\ref{eq:collision-rule})
we find\begin{equation}
\gamma\tau_{n}=\dot{\Psi}_{n}-\dot{\Xi}_{n+1}\label{eq:Anna's-formula}\end{equation}
 which may be rewritten as \begin{equation}
2\gamma\tau_{n}+\dot{Y}_{n+1}-\dot{Y}_{n}=\dot{X}_{n}+\dot{X}_{n+1}.\label{eq:consequence-of-Annas-formula}\end{equation}
 Because the l.h.s. tends to zero as $n\rightarrow\infty$, and $\dot{X}_{n}\ge0$,
then we have $\lim_{n\rightarrow\infty}\dot{X}_{n}=0$. But, considering
that $X_{n}^{(k)}$ are bounded quantities, for $t\in[t_{n},t_{n+1})$
we may write \[
x(t)=\dot{X}_{n}(t-t_{n})+O(\tau_{n}^{2})\]
 and \[
\dot{x}(t)=\dot{X}_{n}+O(\tau_{n}).\]
 Therefore we have: \begin{equation}
\lim_{t\rightarrow t_{\infty}}\dot{x}(t)=\lim_{t\rightarrow t_{\infty}}x(t)=0.\label{eq:X-Xdot-Infty}\end{equation}
 Furthermore, recalling that $\ddot{y}$ is bounded, we have that
$\dot{Y}_{n+1}-\dot{Y}_{n}=O(\tau_{n})$. Observing that $\dot{X}_{n}$
is never negative, from (\ref{eq:consequence-of-Annas-formula}) follows
\begin{equation}
\dot{X}_{n}=O(\tau_{n}).\label{eq:XdotOtau}\end{equation}
 Having assumed that $\sum_{n}\tau_{n}$ converges, we must conclude
that $\sum_{n}\dot{X}_{n}$ converges as well.

The expressions for $X_{n}^{(k)}$ and $Y_{n}^{(k)}$, found by differentiating
repeatedly the equations of motion (\ref{eq:dimensionless-eq-motion}),
may all be written as continuous functions of $\dot{X}_{n},$ $Y_{n}$,
$\dot{Y}_{n}$, which have a limit. Then there also exist $\lim_{n\rightarrow\infty}X_{n}^{(k)}=X_{\infty}^{(k)}$
and $\lim_{n\rightarrow\infty}Y_{n}^{(k)}=Y_{\infty}^{(k)}$.

Higher derivatives of $x$ and $y$ are generally discontinuous in
$t_{n}$, but their jumps are all amenable to the jumps of $\dot{x}$.
To prove this fact we observe that the collision rule (\ref{eq:collision-rule-x1})
may be written as\begin{equation}
[\dot{x}(t_{n})]=2\dot{X}_{n}\label{eq:collision-rule-jump}\end{equation}
 where $[\dot{x}(t_{n})]=\dot{x}(t_{n}^{+})-\dot{x}(t_{n}^{-})$ is
the jump of the function $\dot{x}$ around $t_{n}$. Then, from the
first of equations (\ref{eq:dimensionless-eq-motion}), recalling
that $x\in C^{0}[0,t_{\infty})$ and $y\in C^{1}[0,t_{\infty})$,
it follows that \[
[\ddot{x}(t_{n})]=-2\mu\dot{X}_{n}\]
 and from $\ddot{x}+\ddot{y}=-2\gamma$ we have \[
[\ddot{y}(t_{n})]=2\mu\dot{X}_{n}.\]
 Analogously, differentiating equations (\ref{eq:dimensionless-eq-motion})
with respect to time, and using $\dddot{x}+\dddot{y}=0$ we may evaluate
the jumps of the third derivatives:\[
[\dddot{x}(t_{n})]=-[\dddot{y}(t_{n})]=(4\mu^{2}-1)\dot{X}_{n}.\]
 Iterating the procedure we find that there exists a sequence of real
numbers $\bar{c}_{k}$ such that\begin{equation}
[x^{(k)}(t_{n})]=-[y^{(k)}(t_{n})]=\bar{c}_{k}\dot{X}_{n}.\label{eq:k-derivative_jump}\end{equation}

Next we observe that \[
Y_{\infty}^{(k)}-Y_{n}^{(k)}=\sum_{h=n}^{\infty}\left(Y_{h+1}^{(k)}-Y_{h}^{(k)}\right)=\int_{t_{n}}^{t_{\infty}}y^{(k+1)}(t)\, dt+\sum_{h=n+1}^{\infty}[y^{(k)}(t_{h})].\]
 Using (\ref{eq:x-y-boundedness}) and (\ref{eq:k-derivative_jump})
we have\begin{equation}
\left|Y_{\infty}^{(k)}-Y_{n}^{(k)}\right|\le c_{k+1}T_{n}+\bar{c}_{k}V_{n}\label{eq:Yk-estimate}\end{equation}
 where we have defined the time left to collapse\begin{equation}
T_{n}=t_{\infty}-t_{n}\label{eq:Tn-definition}\end{equation}
 and the residual sum\begin{equation}
V_{n}=\sum_{h=n+1}^{\infty}\dot{X}_{h}.\label{eq:residual-sumXdot}\end{equation}
 Taking $t_{n}\le t\le t_{n+1}$, from (\ref{eq:x-y-boundedness})
and (\ref{eq:Yk-estimate}) evaluated at $n+1$, it follows \begin{equation}
\left|Y_{\infty}^{(k)}-y^{(k)}(t)\right|\le c_{k+1}(t_{\infty}-t)+\bar{c}_{k}V_{n}.\label{eq:Ykinfty-ykt}\end{equation}
 Observing that\[
Y_{\infty}^{(k)}-y^{(k)}(t)=\int_{t}^{t_{\infty}}y^{(k+1)}(s)ds+\sum_{i=n+1}^{\infty}[y^{(k)}(t_{i})]\]
 and using (\ref{eq:Ykinfty-ykt}) for $k+1$ we have\[
Y_{\infty}^{(k)}-y^{(k)}(t)=\int_{t}^{t_{\infty}}\left(Y_{\infty}^{(k+1)}+O(t_{\infty}-s)\right)ds+O(V_{n})=Y_{\infty}^{(k+1)}(t_{\infty}-t)+O\left((t_{\infty}-t)^{2}\right)+O(V_{n})\]
 hence\begin{equation}
y^{(k)}(t)=Y_{\infty}^{(k)}+Y_{\infty}^{(k+1)}(t-t_{\infty})+O((t-t_{\infty})^{2})+O(V_{n}).\label{eq:yk-expansion}\end{equation}
 Choosing $k=2$, by repeated integration of (\ref{eq:yk-expansion}),
exploiting the continuity of $y$ and $\dot{y}$ up to $t_{\infty}$,
we obtain\begin{equation}
\dot{y}(t)=\dot{Y}_{\infty}+\ddot{Y}_{\infty}(t-t_{\infty})+\frac{1}{2}\dddot{Y}_{\infty}(t-t_{\infty})^{2}+O((t-t_{\infty})^{3})+O((t_{\infty}-t)V_{n})\label{eq:ydot-expansion}\end{equation}
 and \begin{multline}
y(t)=Y_{\infty}+\dot{Y}_{\infty}(t-t_{\infty})+\frac{1}{2}\ddot{Y}_{\infty}(t-t_{\infty})^{2}+\\
+\frac{1}{6}\dddot{Y}_{\infty}(t-t_{\infty})^{3}+O((t-t_{\infty})^{4})+O((t_{\infty}-t)^{2}V_{n})\label{eq:y-expansion}\end{multline}

Using (\ref{eq:ydot-expansion}) and (\ref{eq:y-expansion}) in the
first of equations (\ref{eq:dimensionless-eq-motion}) we find that
the motion of the lower point mass, under the assumption of inelastic
collapse, would be described by the following equation\begin{equation}
\ddot{x}+\mu\dot{x}+\frac{1}{2}x=P-Q(t_{\infty}-t)+R(t_{\infty}-t)^{2}+O((t_{\infty}-t)^{3})+O((t_{\infty}-t)V_{n})\label{eq:collapse-equation}\end{equation}
 where \begin{eqnarray}
P & = & -\gamma-\frac{1}{2}+\frac{Y_{\infty}}{2}+\mu\dot{Y}_{\infty},\nonumber \\
Q & = & \frac{\dot{Y}_{\infty}}{2}+\mu\ddot{Y}_{\infty},\label{eq:PQR}\\
R & = & \frac{\ddot{Y}_{\infty}}{4}+\mu\frac{\dddot{Y}_{\infty}}{2}.\nonumber \end{eqnarray}
 Because $P,$ $Q,$ $R$ are numbers, the motion of the lower point
mass is now decoupled from that of the upper point mass, and equation
(\ref{eq:collapse-equation}) is closed. Furthermore, we may evaluate
the equations of motion (\ref{eq:dimensionless-eq-motion}) and their
first two time derivatives as $t\rightarrow t_{n}^{+}$. Using (\ref{eq:ydot-expansion}),
(\ref{eq:y-expansion}) and (\ref{eq:yk-expansion}) for $k=2,3$
in order to eliminate $y$ and its derivatives, observing that $\dot{X}_{n}=O(T_{n})$
because of (\ref{eq:XdotOtau}), we find the following asymptotic
relationships \begin{equation}
\left\{ \begin{array}{l}
\ddot{X}_{n}=P-\mu\dot{X}_{n}-QT_{n}+RT_{n}^{2}+O(T_{n}^{3})+O(T_{n}V_{n})\\
\dddot{X}_{n}=Q-\mu P+(\mu Q-2R)T_{n}+O(T_{n}^{2})+O(V_{n})\\
X_{n}^{\mathrm{(4})}=(\mu^{2}-\frac{1}{2})P-\mu Q+2R+O(T_{n})+O(V_{n})\end{array}\right..\label{eq:X-derivatives}\end{equation}

\subsection{\label{sub:NoCollapse}Absence of Inelastic Collapse}

In the interval between two consecutive impacts, the motion of the
point masses is described by smooth functions. Thereby, there exist
two numbers $\zeta_{n}$ and $\eta_{n}$, with $t_{n}<\zeta_{n},\eta_{n}<t_{n+1}$,
such that the following expressions hold for $t_{n}\le t\le t_{n+1}$\begin{equation}
x(t_{n+1})=\dot{X}_{n}\tau_{n}+\ddot{X}_{n}\frac{\tau_{n}^{2}}{2}+\dddot{x}(\zeta_{n})\frac{\tau_{n}^{3}}{6}\label{eq:x-exact-expansion}\end{equation}
 and \begin{equation}
\dot{x}(t_{n+1})=\dot{X}_{n}+\ddot{X}_{n}\tau_{n}+\dddot{x}(\eta_{n})\frac{\tau_{n}^{2}}{2}.\label{eq:xdot-exact-expansion}\end{equation}
 Since $x(t_{n+1})=0$, from (\ref{eq:x-exact-expansion}) we find
\begin{equation}
\ddot{X}_{n}\tau_{n}=-2\dot{X}_{n}-\dddot{x}(\zeta_{n})\frac{\tau_{n}^{2}}{3}.\label{eq:xddot-tau}\end{equation}
 Inserting this expression in (\ref{eq:xdot-exact-expansion}) and
applying the collision rule (\ref{eq:collision-rule-map}), we obtain
the following map for the speed of the rebounds\begin{equation}
\dot{X}_{n+1}=\dot{X}_{n}-\Delta_{n}\tau_{n}^{2}\label{eq:xdot_map}\end{equation}
 where \begin{equation}
\Delta_{n}=\left(\dddot{x}(\eta_{n})/2-\dddot{x}(\zeta_{n})/3\right).\label{eq:Delta-n}\end{equation}

With the help of the map (\ref{eq:xdot_map}), we shall examine the
dynamics generated by equation (\ref{eq:collapse-equation}) separately
for the three cases $P\neq0$; $P=0,$ $Q\neq0$; and $P=0,$ $Q=0$;
that is $\ddot{X}_{\infty}\neq0$; $\ddot{X}_{\infty}=0,$ $\dddot{X}_{\infty}\neq0$;
and $\ddot{X}_{\infty}=0,$ $\dddot{X}_{\infty}=0$. They are exhaustive
of all possible asymptotic behaviors of the ball because, from the
definitions (\ref{eq:PQR}), $P=0,$ $Q=0$ implies $R=-\gamma/2$.
We shall see that they all are in contradiction with the hypothesis
of inelastic collapse.

\subsubsection{The case $\ddot{X}_{\infty}\neq0$ }

Recalling that $\dddot{x}$ is a bounded function, and that $\lim_{n\rightarrow\infty}\tau_{n}=0$,
(\ref{eq:xddot-tau}) yields\begin{equation}
\tau_{n}=O(\dot{X}_{n})\label{eq:tauOXdot}\end{equation}
 and also\[
\ddot{X}_{\infty}=\lim_{n\rightarrow\infty}\left(-2\frac{\dot{X}_{n}}{\tau_{n}}+O(\tau_{n}^{2})\right)<0.\]
 This implies $\dot{X}_{n}>0$, $\ddot{X}_{n}<0$ and $\tau_{n}\simeq\dot{X}_{n}$
for large $n$. Then (\ref{eq:xdot_map}) gives \[
\dot{X}_{n+1}=\dot{X}_{n}+O(\dot{X}_{n}^{2}).\]
 From Lemma (\ref{lem:Giorgio's-2}) follows that $\sum_{n}\dot{X}_{n}=+\infty$,
then $\sum_{n}\tau_{n}$ diverges as well. Thus the inelastic collapse
can not occur in this case.

\subsubsection{The case $\ddot{X}_{\infty}=0$ and $\dddot{X}_{\infty}\neq0$}

Equations (\ref{eq:x-exact-expansion}) through (\ref{eq:Delta-n}),
are still valid because they do not depend on the asymptotic behavior
of $\ddot{X}_{n}$ and $\dddot{X}_{n}$. In particular, in the limit
$n\rightarrow\infty$, since $\dddot{x}(\zeta_{n}),\dddot{x}(\eta_{n})=\dddot{X}_{n}+O(\tau_{n})$
in the definition (\ref{eq:Delta-n}), we have $\Delta_{n}\rightarrow\dddot{X}_{\infty}/6$.
If we assume $\dddot{X}_{\infty}<0$, the map (\ref{eq:xdot_map})
does not allow $\dot{X}_{n}\rightarrow0$, which is inconsistent with
(\ref{eq:xddot-tau}) and $\tau_{n}\rightarrow0$. Then $\dddot{X}_{\infty}\neq0$
implies  $\dddot{X}_{n}>0$ for large $n$. Furthermore, $\Delta_{n}>0$,
together with $\tau_{n}>0$ implies $\dot{X}_{n}>0$. In fact, if
$\dot{X}_{n}=0$ for some $n$, the map (\ref{eq:xdot_map}) would
give $\dot{X}_{n+1}<0$ which is in contradiction with the constraint
$x(t)\ge0$ imposed by the presence of the floor.

Using $\dddot{X}_{n},\tau_{n},\dot{X}_{n}>0$, from equation (\ref{eq:xddot-tau}),
we find that\[
\ddot{X}_{n}<0\]
 for all $n$ large enough. Furthermore, the hypothesis $\ddot{X}_{\infty}=0$
implies \begin{equation}
\lim_{n\rightarrow\infty}\frac{\dot{X}_{n}}{\tau_{n}}=0.\label{eq:Xdot-fasterthan-tau}\end{equation}
 Observing that $T_{n}=\sum_{h=n}^{\infty}\tau_{n}$ we also have\begin{equation}
\lim_{n\rightarrow\infty}\frac{\dot{X}_{n}}{T_{n}}=0.\label{eq:Xdot-fasterthan-Tn}\end{equation}
 Using (\ref{eq:xdot_map}) and (\ref{eq:Tn-definition}) we can write
the following map for the pair $(\dot{X}_{n},T_{n})$:\begin{equation}
\left\{ \begin{array}{ccc}
\dot{X}_{n+1} & = & \dot{X}_{n}-\Delta_{n}\tau_{n}^{2}\\
T_{n+1} & = & T_{n}-\tau_{n}\end{array}\right..\label{eq:Tha-Map}\end{equation}
 In order to have an inelastic collapse we need to find at least one
sequence of pairs $(\dot{X}_{n},T_{n})\rightarrow(0,0)$ as $n\rightarrow0$
which satisfies this this map and (\ref{eq:Xdot-fasterthan-Tn}).

Having proved that $\dot{X}_{n},\dddot{X}_{n}>0$ and $\ddot{X}_{n}<0$
for $n$ large enough, recalling that $x(t_{n+1})=0$, the time of
flight computed from (\ref{eq:x-exact-expansion}) is \begin{equation}
\tau_{n}=\frac{3}{2}\frac{|\ddot{X}_{n}|}{\dddot{x}(\zeta_{n})}\left(1-\sqrt{1-\frac{8}{3}\frac{\dddot{x}(\zeta_{n})\dot{X}_{n}}{\ddot{X}_{n}^{2}}}\right).\label{eq:tau3}\end{equation}
 With (\ref{eq:Tha-Map}) and (\ref{eq:tau3}) we compute\begin{equation}
\frac{T_{n}}{\dot{X}_{n}}\frac{\dot{X}_{n}-\dot{X}_{n+1}}{T_{n}-T_{n+1}}=4\frac{\Delta_{n}}{\left|\ddot{X}_{n}\right|}\frac{T_{n}}{\left(1+\sqrt{1-\frac{8}{3}\frac{\dddot{x}_{n}(\zeta_{n})\dot{X}_{n}}{\ddot{X}_{n}^{2}}}\right)}\label{eq:strange-combination}\end{equation}
 From (\ref{eq:X-derivatives}) and (\ref{eq:Xdot-fasterthan-Tn}),
recalling that $P=\ddot{X}_{\infty}=0$, we have $\ddot{X}_{n}\sim-QT_{n}$
and $\dddot{X}_{n}\sim Q$, that is $\Delta_{n}\sim Q/6$. Using these
asymptotic expressions in (\ref{eq:strange-combination}) we find
\begin{equation}
\limsup_{n\rightarrow\infty}\frac{T_{n}}{\dot{X}_{n}}\frac{\dot{X}_{n}-\dot{X}_{n+1}}{T_{n}-T_{n+1}}\le\frac{2}{3}\label{eq:tanalpha-tanbeta}\end{equation}
 which, in turn, implies\begin{equation}
\frac{\dot{X}_{n+1}}{T_{n+1}}>\frac{\dot{X}_{n}}{T_{n}}.\label{eq:contradiction}\end{equation}
 This is in contradiction with (\ref{eq:Xdot-fasterthan-Tn}) .

To conclude the study of this case we observe that the second non-zero
root of (\ref{eq:x-exact-expansion}), namely\begin{equation}
\bar{\tau}_{n}=\frac{3}{2}\frac{|\ddot{X}_{n}|}{\dddot{x}(\bar{\zeta}_{n})}\left(1+\sqrt{1-\frac{8}{3}\frac{\dddot{x}(\bar{\zeta}_{n})\dot{X}_{n}}{\ddot{X}_{n}^{2}}}\right)\label{eq:tau3bar}\end{equation}
 can not be the time of flight for large $n$, because $\ddot{X}_{n}\sim-QT_{n}$
and $\dddot{X}_{n}\sim Q$ would lead to the absurd $\bar{\tau}_{n}>T_{n}$.

\subsubsection{The case $\ddot{X}_{\infty}=0$ and $\dddot{X}_{\infty}=0$}

In this case we have $P=0$, $Q=0$ in (\ref{eq:X-derivatives}) so
that $\ddot{X}_{n}=o(T_{n})$. Looking at (\ref{eq:xddot-tau}) now
we find \begin{equation}
\dot{X}_{n}=o(\tau_{n}T_{n})\label{eq:Xdot-opiccolo}\end{equation}
 which implies that the residual sum (\ref{eq:residual-sumXdot})
is \begin{equation}
V_{n}=o(T_{n}^{2}).\label{eq:Vn-opiccolo}\end{equation}
 From (\ref{eq:PQR}), with the help of the equations of motion (\ref{eq:dimensionless-eq-motion})
and their derivatives, we have $R=-\gamma/2$. Using these relations
back in (\ref{eq:X-derivatives}) we find\begin{equation}
\left\{ \begin{array}{l}
\ddot{X}_{n}=-\frac{1}{2}\gamma T_{n}^{2}+o(T_{n}^{2})\\
\dddot{X}_{n}=\gamma T_{n}+O(T_{n}^{2})\\
X_{n}^{(4)}=-\gamma+O(T_{n})\end{array}\right..\label{eq:asympt-4ord}\end{equation}
 The first of these equalities show that $\ddot{X}_{n}<0$ for large
$n$. Then $\dot{X}_{n}>0$ for equally large $n$. In fact, $\dot{X}_{n}=0$
and $\ddot{X}_{n}<0$ imply $x(t)<0$ for $t$ close to $t_{n}$,
which is incompatible with the floor constraint $x(t)\ge0$.

Evaluating $x(t_{n+1})$ with a power series expansion around $t_{n}$,
we have \begin{equation}
\dot{X}_{n}=-\left(\frac{\ddot{X}_{n}}{2}\tau_{n}+\frac{\dddot{X}_{n}}{6}\tau_{n}^{2}+\frac{X_{n}^{(4)}}{24}\tau_{n}^{3}\right)+O(\tau_{n}^{4}).\label{eq:taylor-x-4ord}\end{equation}
 In the same way, a power series for $\dot{x}(t_{n+1})$ together
with the collision rule (\ref{eq:collision-rule-map}) gives\begin{equation}
\dot{X}_{n+1}=-\dot{X}_{n}-\ddot{X}_{n}\tau_{n}-\frac{1}{2}\dddot{X}_{n}\tau_{n}^{2}-\frac{1}{6}X_{n}^{(4)}\tau_{n}^{3}+O(\tau_{n}^{4}).\label{eq:taylor-xdot-4ord}\end{equation}
 We use (\ref{eq:taylor-x-4ord}) in (\ref{eq:taylor-xdot-4ord})
in order to eliminate $\dot{X}_{n}$ and $\dot{X}_{n+1}$, we substitute
the higher derivatives with the expressions (\ref{eq:asympt-4ord}),
and we write $\tau_{n}=\alpha_{n}T_{n}$ for some $\alpha_{n}\in(0,1)$.
After some algebra, this yields the following implicit map\begin{equation}
\left(1-\alpha_{n}\right)^{3}\left(6\alpha_{n+1}-4\alpha_{n+1}^{2}+\alpha_{n+1}^{3}\right)=6\alpha_{n}-8\alpha_{n}^{2}+3\alpha_{n}^{3}+o(1)\label{eq:alpha-map}\end{equation}
 where the term $o(1)$ goes to zero as $n\rightarrow\infty$. We
prove in Lemma (\ref{lem:alpha-map}) that if the sequence $\left\{ \alpha_{n}\right\} $
stays in $(0,1)$, then it converges to zero, that is $\tau_{n}=o(T_{n})$.
Of course, $\alpha_{n}>1$ for some $n$ would lead to $\tau_{n}>T_{n}$,
which is an absurd.

Having proved that, for large $n$, it is $\dot{X}_{n}>0,$ $\ddot{X}_{n}<0,$
$\dddot{X}_{n}>0$, it follows that expressions (\ref{eq:Xdot-fasterthan-Tn})
through (\ref{eq:strange-combination}) still hold in this case. Using
(\ref{eq:asympt-4ord}) in (\ref{eq:taylor-x-4ord}) we find $\dot{X}_{n}\sim\frac{\gamma}{4}T_{n}^{2}\tau_{n}$.
Then, recalling that $\dddot{x}(\zeta_{n})=\dddot{X}_{n}+O(\tau_{n})$,
we find \[
\frac{\dddot{x}(\zeta_{n})\dot{X}_{n}}{\ddot{X}_{n}^{2}}\sim\frac{\tau_{n}}{T_{n}}\]
 and we conclude that the square root in (\ref{eq:strange-combination})
approaches one for large $n$. Using $\Delta_{n}\sim\dddot{X}_{n}/6$
and (\ref{eq:asympt-4ord}) in (\ref{eq:strange-combination}), in
this case gives\[
\lim_{n\rightarrow\infty}\frac{T_{n}}{\dot{X}_{n}}\frac{\dot{X}_{n}-\dot{X}_{n+1}}{T_{n}-T_{n+1}}=\frac{2}{3}.\]
 Thus we have again \[
\frac{\dot{X}_{n+1}}{T_{n+1}}>\frac{\dot{X}_{n}}{T_{n}}\]
 which contradicts (\ref{eq:Xdot-fasterthan-Tn}). Finally, we observe
that, also in this case, (\ref{eq:tau3bar}) gives $\bar{\tau}_{n}>T_{n}$.

\section{\label{sec:Asymptotic-State}Asymptotic State}

We now carry on to prove Theorem (\ref{thm:static-equilibrium}).
If the asymptotic dynamics is sticky, then equation (\ref{eq:y-eq-sticky})
holds for all times greater than the last attachment time $t_{s}.$
It is straightforward to verify that as $t\rightarrow+\infty$ the
system reaches the equilibrium (\ref{eq:equilibrium}) .

If the asymptotic dynamics is not sticky, then a rigorous proof is
not trivial. A useful information, more or less implicit in the calculations
of the previous section, is that\begin{equation}
\dot{X}_{n}=O(\tau_{n-1}).\label{eq:Xdot-is-Otau}\end{equation}
 To see that this equality holds, we remark that there exists a number
$\eta_{n-1}\in(t_{n-1},t_{n})$ such that\[
-\dot{X}_{n}=\dot{X}_{n-1}+\ddot{x}(\eta_{n-1})\tau_{n-1}\]
 which implies (\ref{eq:Xdot-is-Otau}), since $\dot{X}_{n}$ is not
negative for any $n$.

We split the proof in two lemmas, and a final theorem. Both lemmas,
ideally, are nothing more than a Lyapunov fixed point theorem, but
they have to deal with technicalities arising from the presence of
the impacts.

\begin{lem}
\label{lem:xidot(t)-goes-to-0}If the asymptotic dynamics is not sticky,
then $\lim_{t\to\infty}\dot{\xi}(t)=0.$ 
\end{lem}
\begin{proof}
We assume by contradiction that \begin{equation}
\limsup_{t\to\infty}\left|\dot{\xi}(t)\right|>0.\label{eq:abs1}\end{equation}
 If \eqref{eq:abs1} holds, having excluded the presence of inelastic
collapse, it is possible to find a monotonically growing sequence
of times $s_{k}\to\infty$, with $s_{k+1}-s_{k}>2$ and $\limsup_{k\rightarrow\infty}\left|\dot{\xi}(s_{k})\right|>0$.
Because $\xi$ and all its derivatives are bounded functions, recalling
that $\dot{E}(t)=-2\mu\dot{\xi}^{2}$ (see eq. \ref{eq:xi-psi}) we
see that there exist positive $\delta$ and $C$ such that \begin{equation}
\dot{E}(s_{k})\le-\delta\label{eq:stimaderivata}\end{equation}
 and\begin{equation}
\left|\ddot{E}(t)\right|\le C\label{eq:stimaderseconda}\end{equation}
 for all $k$ and for all $t\in\cup_{n=0}^{\infty}(t_{n},t_{n+1})$.
We claim that there is a $\hat{t}$ such that $\left|\hat{t}-s_{k}\right|\ge\delta/(16C)$
and $\dot{E}(t)\le-\delta/2$ for any $t$ that is not a time of impact
in the interval having $\hat{t}$ and $s_{k}$ as extremes. Let $t_{n}$,
$t_{n+1}$ be the times of consecutive impacts that bracket $s_{k}$:
to prove the claim we have to consider two different cases:

1. If $t_{n+1}-t_{n}\ge\delta/(8C)$, then there exists $\hat{t}\in[t_{n},t_{n+1}]$
such that \begin{equation}
\left|\hat{t}-s_{k}\right|=\frac{\delta}{16C}.\label{eq:delta16}\end{equation}
 From \eqref{eq:stimaderseconda}, for any $t$ between $\hat{t}$
and $s_{k}$ we have \[
\frac{\dot{E}(t)-\dot{E}(s_{k})}{t-s_{k}}=\ddot{E}(\zeta)\le C\]
 for some $\zeta$ between $t$ and $s_{k}$. Observing that $\left|t-s_{k}\right|\le\delta/(16C)$,
from \eqref{eq:stimaderivata} we get\[
\dot{E}(t)\le\dot{E}(s_{k})+\frac{\delta}{16}<-\frac{\delta}{2}\]

2. If \begin{equation}
t_{n+1}-t_{n}<\frac{\delta}{8C}\label{eq:tau-minoraz}\end{equation}
 then let $m$ be the integer such that \begin{equation}
\left\{ \begin{array}{ccc}
t_{n+m+1}-t_{n} & > & \delta/(4C)\\
t_{n+m}-t_{n} & \le & \delta/(4C)\end{array}\right..\label{eq:intervalli}\end{equation}
 There exists $\hat{t}\in[t_{n+m},t_{n+m+1}]$ such that \begin{equation}
\hat{t}-t_{n}=\frac{\delta}{4C}.\label{eq:int-hat}\end{equation}
 Recalling that $t_{n}\le s_{k}\le t_{n+1}$ then \[
\hat{t}-s_{k}\ge\hat{t}-t_{n+1}=(\hat{t}-t_{n})-(t_{n+1}-t_{n})\]
 and from (\ref{eq:tau-minoraz}) and (\ref{eq:intervalli}) we get\begin{equation}
\hat{t}-s_{k}>\frac{\delta}{8C}\label{eq:that-sk-magg}\end{equation}
 From (\ref{eq:xi-psi}) we have $\dot{\xi}^{2}=\left(\dot{x}^{2}-2\dot{x}\dot{y}+\dot{y}^{2}\right)/4$
and from (\ref{eq:collision-rule-jump}) follows that the jump of
$\dot{\xi}^{2}$ across an impact is \begin{equation}
[\dot{\xi}^{2}(t_{n})]=-\dot{X}_{n}\dot{y}(t_{n}).\label{eq:xidot2-jump}\end{equation}
 Recalling that $\dot{E}(t)=-2\mu\dot{\xi}^{2}$ and that $\dot{X}_{n}=O(\tau_{n-1})$,
we have can choose the constant $C$ in (\ref{eq:stimaderseconda})
in such a way that \begin{equation}
[\dot{E}(t_{n})]\le C\tau_{n-1}\label{eq:jump}\end{equation}
 for all $n$. Moreover, for $t\in[s_{k},\hat{t}]$ such that $t_{n+s}<t<t_{n+s+1}$
for some $s\le m$, \begin{multline}
\dot{E}(t)-\dot{E}(s_{k})=\left(\dot{E}(t)-\dot{E}(t_{n+s}^{+})\right)+\left(\dot{E}(t_{n+s}^{-})-\dot{E}(t_{n+s-1}^{+})\right)+\cdots\label{eq:somma1}\\
\cdots+\left(\dot{E}(t_{n+1}^{-})-\dot{E}(s_{k})\right)+\sum_{i=1}^{s}[\dot{E}(t_{n+i})].\end{multline}
 Using (\ref{eq:stimaderseconda}) and (\ref{eq:jump}) we obtain\[
\dot{E}(t)-\dot{E}(s_{k})\le C\left(t-s_{k}\right)+C\left(t_{n+m}-t_{n}\right)\le2C\left(\hat{t}-t_{n}\right)=\frac{\delta}{2}\]
 and, from (\ref{eq:stimaderivata}) \[
\dot{E}(t)<-\frac{\delta}{2}.\]

The claim is proved, and for any $k$ there exists an interval not
shorter than $\delta/(16C)$ having $s_{k}$ as an extremum. During
each of those intervals the power dissipated is at least $-\delta/2$,
which gives an infinite dissipation of the mechanical energy, in contradiction
with the fact that $E(t)$ has a lower bound. Then (\ref{eq:abs1})
is false, and we must conclude that $\lim_{t\to\infty}\dot{\xi}(t)=0.$ 
\end{proof}
\begin{lem}
\label{lem:xi(t)-goes-to-0}If the asymptotic dynamics is not sticky,
then $\lim_{t\to\infty}\xi(t)=-\gamma.$ 
\end{lem}
\begin{proof}
The proof is identical to that of the previous lemma, except that,
in place of the mechanical energy, we use an \emph{ad hoc} Lyapunov
function $\mathcal{L}$ defined as \[
\mathcal{L}(t)=\frac{1}{2}[(\dot{\xi}+2\mu(\xi+\gamma))^{2}+(\dot{\psi}+2\mu(\xi+\gamma))^{2}+\xi^{2}]+\gamma\psi.\]
 Because $\psi,$ $\xi$ and all their derivatives are bounded, $\mathcal{L}$
and all its derivatives are bounded as well. We observe that $\mathcal{L}$
is a continuous function of time, even across the times of impact.
Furthermore we find\begin{equation}
\dot{\mathcal{L}}(t)=-2\mu(\xi(t)+\gamma)^{2}+O(\dot{\xi}).\label{eq:Vdot}\end{equation}
 We assume by contradiction that $\limsup_{t\to\infty}\left|\xi(t)+\gamma\right|>0$.
Proceeding as in the previous lemma, and exploiting the fact that
$\lim_{t\rightarrow0}\dot{\xi}(t)=0$ makes (\ref{eq:Vdot}) negative
if $\left|\xi(t)-\gamma\right|$ is bounded from below from zero,
we conclude that $\lim_{t\rightarrow\infty}\mathcal{L}(t)=-\infty$,
which contradicts the boundedness of $\mathcal{L}$. Therefore $\lim_{t\to\infty}\left(\xi(t)+\gamma\right)=0$. 
\end{proof}

\subsection{Proof of Theorem (\ref{thm:static-equilibrium})}

From the collision condition (\ref{eq:collision-condition-psi-xi})
and lemma (\ref{lem:xi(t)-goes-to-0}) it follows\[
\lim_{n\rightarrow\infty}\psi(t_{n})=-\gamma.\]
 From the collision rule (\ref{eq:collision-rule}) and lemma (\ref{lem:xidot(t)-goes-to-0})
it follows \[
\lim_{n\rightarrow\infty}\dot{\Psi}_{n}=0\]
and from (\ref{eq:tot-mech-energy}) we deduce that $\lim_{n\rightarrow\infty}E(t_{n}^{+})=-\gamma/2.$
Since the energy is a continuous, decreasing function, then \[
\lim_{t\rightarrow\infty}E(t)=-\frac{\gamma}{2}.\]
 The only state with this energy is the state of static equilibrium
(\ref{eq:equilibrium}).

\subsection{Proof of the Main Theorem (\ref{thm:main_thm})}

\begin{proof}
Theorem (\ref{thm:static-equilibrium}), together with (\ref{eq:Anna's-formula}),
yields\begin{equation}
\lim_{n\to\infty}\tau_{n}=0\label{eq:limtau}\end{equation}
 By Theorem (\ref{thm:static-equilibrium}) we have also \[
\lim_{t\to\infty}y(t)=1-2\gamma,\quad\lim_{t\to\infty}x(t)=\lim_{t\to\infty}\dot{x}(t)=\lim_{t\to\infty}\dot{y}(t)=0\]
 From the equations of motion (6) and their derivatives we can easily
compute \[
\lim_{n\to\infty}\stackrel{..}{X}_{n}=-2\gamma\quad,\quad\lim_{n\to\infty}\stackrel{...}{X}_{n}=2\mu\gamma\]
 We follow here the steps of section \ref{sub:NoCollapse} in order
to write a map for $\dot{X}_{n}$. From (\ref{eq:x-exact-expansion})
we get $\tau_{n}$ as in (\ref{eq:tau3}) (the other non-zero root
of (\ref{eq:x-exact-expansion}) would contradict (\ref{eq:limtau})).
We note that for $\zeta_{n}\in\left(t_{n},t_{n+1}\right)$, $\stackrel{...}{x}(\zeta_{n})=\stackrel{...}{X}_{n}+O(\tau_{n})$,
so\begin{equation}
\lim_{n\to\infty}\stackrel{...}{x}(\zeta_{n})=2\mu\gamma\label{eq:limitex3dot}\end{equation}
 Using (\ref{eq:xdot-exact-expansion}) and (\ref{eq:xddot-tau})
we can write\begin{equation}
\dot{X}_{n+1}=\dot{X}_{n}-\Delta_{n}\tau_{n}^{2}\label{eq:premappa}\end{equation}
 with $\Delta_{n}$ defined in (\ref{eq:Delta-n}). From (\ref{eq:limitex3dot})
we have\begin{equation}
\lim_{n\to\infty}\Delta_{n}=\frac{\mu\gamma}{3}\label{eq:delta}\end{equation}
 Then, expanding the square-root in (\ref{eq:tau3}), from (\ref{eq:premappa})
we get the map\begin{equation}
\dot{X}_{n+1}=\dot{X}_{n}-\alpha_{n}\ \dot{X}_{n}^{2}+O(\dot{X}_{n}^{3})\label{eq:mappa}\end{equation}
 with $\alpha_{n}=4\ \Delta_{n}/\stackrel{..}{X_{n}^{2}}$. Since
$\lim_{n\to\infty}\alpha_{n}=\mu/(3\gamma)$ we can rewrite (\ref{eq:mappa})
as\begin{equation}
\dot{X}_{n+1}=\left(1-\frac{\mu}{3\ \gamma}\dot{X}_{n}\right)\dot{X}_{n}+o(\dot{X}_{n}^{2})\label{eq:restituzione}\end{equation}
 Applying Lemma (\ref{lem:Giorgio's}) to (\ref{eq:restituzione})
we get\[
\dot{X}_{n}\sim\frac{3\gamma}{\mu}\ \frac{1}{n}.\]
 Dividing both sides of (\ref{eq:xddot-tau}) by $\tau_{n}$ we get\[
\frac{\dot{X}_{n}}{\tau_{n}}=-\frac{\stackrel{..}{X}_{n}}{2}+\stackrel{...}{x}(\zeta_{n})\tau_{n}\]
 then\[
\lim_{n\to\infty}\frac{\dot{X}_{n}}{\tau_{n}}=\gamma\]
 which proves the thesis. 
\end{proof}

\section{\label{sec:ZeroMeasure}Non-Genericity of Zero-Velocity Contacts}

In this section we give a proof of Theorem (\ref{thm:contact-non-generic}).
The state of our idealized ball is a vector $\mathbf{u}\in\mathbb{R}^{4}$.
Without loss of generality, we may assume that the first two components
of $\mathbf{u}$ are the position and the velocity of the lower point
mass. We define the hyperplane $\mathcal{C}$ of contact configurations
as\begin{equation}
\mathcal{C}=\left\{ \mathbf{u}\in\mathbb{R}^{4}\left|\mathbf{u}\cdot\mathbf{e}_{1}=0\right.\right\} \label{eq:C}\end{equation}
 and the plane $\mathcal{S}$ of contact configurations with zero
velocity\begin{equation}
\mathcal{S}=\left\{ \mathbf{u}\in\mathcal{C}\left|\mathbf{u}\cdot\mathbf{e}_{2}=0\right.\right\} .\label{eq:S}\end{equation}

We observe that the vector field $\chi$ that defines the equations
of motion (\ref{eq:dimensionless-eq-motion}), on the hyperplane $\mathcal{C}$
is \[
\chi(0,\dot{x},y,\dot{y})=\left(\begin{array}{c}
\dot{x}\\
-\mu\dot{x}-\gamma-\frac{1}{2}+\frac{1}{2}y+\mu\dot{y}\\
\dot{y}\\
-\mu\dot{y}-\frac{1}{2}y-\gamma+\frac{1}{2}+\mu\dot{x}\end{array}\right).\]
 It follows that $\chi(\mathbf{u})$ is transverse to $\mathcal{C}$
if and only if the second component of $\mathbf{u}$ is non zero.
That is, $\chi$ is transverse to $\mathcal{C}$ everywhere but in
$\mathcal{S}$.

To prove that the set of initial conditions generating orbits that
reach $\mathcal{S}$ in a finite time has zero measure, we need to
define the backward return map of the flow on $\mathcal{C}$. This
requires some care because the points of $\mathcal{S}$ are contiguous
to points that cannot be reached by the dynamics.

The time evolution of state vectors in the interval $(t_{n},t_{n+1})$
is determined by the flow $\varphi:\mathbb{R}\times\mathbb{R}^{4}\rightarrow\mathbb{R}^{4}$
of the equations of motion (\ref{eq:dimensionless-eq-motion}). We
may define an inverse time-of-flight function implicitly as the largest
negative number $\tau^{-}$such that \begin{equation}
\varphi(\tau^{-},\mathbf{u})\cdot\mathbf{e}_{1}=0\label{eq:implicit-tau-minus}\end{equation}
 for any $\mathbf{u}\in\mathcal{C}^{-}$, where $\mathcal{C}^{-}$
is the subset of vectors of $\mathcal{C}$ with $\dot{x}\le0$. Some
care is needed in order to define $\tau^{-}$ for the states belonging
to $\mathcal{S}$. We observe that if $\dot{x}=0$ and $\ddot{x}>0$
we have a grazing event, and definition (\ref{eq:implicit-tau-minus})
is applicable. If $\ddot{x}=0$ then we need to evaluate the force
(\ref{eq:force-on-the-floor}) exherted by the ball on the floor:
if $F(\mathbf{u})=0$ and $\dot{F}(\mathbf{u})<0$ then the state
$\mathbf{u}$ is the beginning of a sticky event, and definition (\ref{eq:implicit-tau-minus})
is again applicable; for all other states on $\mathcal{S}$ a sticky
event has already begun, and we define $\tau^{-}=0$.

We observe that the backward-in-time evolution of a sticky state does
not leave $\mathcal{S}$ as long as $\tau^{-}=0$. On the other hand,
any state with $\tau^{-}\neq0$ leaves $\mathcal{C}$ following the
flow $\varphi$ backward in time. Then, for these states we introduce
the transformation $\mathcal{T}^{-}:\mathcal{C}^{+}\rightarrow\mathcal{C}^{+}$
defined as \begin{equation}
\mathcal{T}^{-}(\mathbf{u})=\varphi(\tau^{-},R\mathbf{u})\label{eq:transformation-Tminus}\end{equation}
where $R$ is the reflection matrix that changes sign to the second
component of $\mathbf{u}$. A sketch of this process is given in figure
\begin{figure}
\includegraphics[width=0.8\columnwidth]{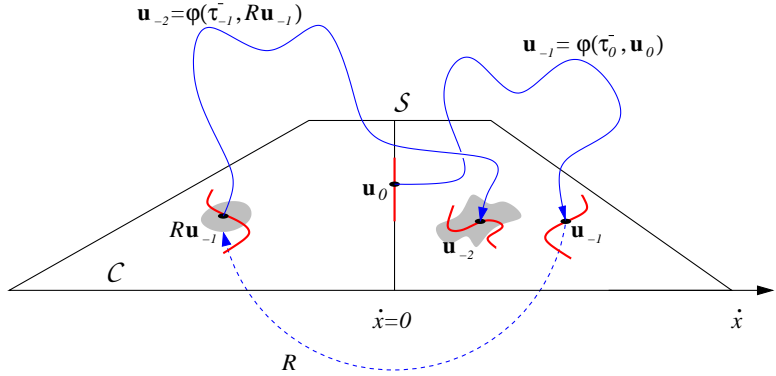}

\caption{\label{fig:zero-measure}In proximity of a non-sticky state $\mathbf{u}_{0}$
that evolves backward in time into a state $\mathbf{u}_{-1}\notin\mathcal{S}$,
a subset of $\mathcal{S}$ (the thick segment) is mapped into a subset
of $\mathcal{C}$ of zero measure (the thick wiggly line). Furthermore,
if $\mathbf{u}_{-1}$ evolves backward into $\mathbf{u}_{-2}\notin\mathcal{S}$
then $\mathcal{T}^{-}$ is a diffeomorphism between a neighborhood
of $\mathbf{u}_{-1}$ and a neighborhood of $\mathbf{u}_{-2}$. }
\end{figure}
 \ref{fig:zero-measure}.

Let us assume that a given pair $(\tau_{0}^{-},\mathbf{u}_{0})$ satisfies
(\ref{eq:implicit-tau-minus}) with $\tau_{0}^{-}\neq0$. If $\varphi(\tau_{0}^{-},\mathbf{u}_{0})\notin\mathcal{S}$,
then \[
0\neq\left.\frac{\partial}{\partial\tau^{-}}\left(\varphi(\tau^{-},\mathbf{u})\cdot\mathbf{e}_{1}\right)\right|_{\tau_{0}^{-},\mathbf{u}_{0}}=\chi\left(\varphi(\tau_{0}^{-},\mathbf{u}_{0})\right)\cdot\mathbf{e}_{1}\]
 and the implicit function theorem yields a smooth mapping $\tau^{-}$
in an open neighborhood of $(\tau_{0}^{-},\mathbf{u}_{0})$. However
this map need not coincide with our inverse time of flight in the
whole neighborhood, since it depends on the free flow $\varphi$ which
does not take into account the presence of the floor. To overcome
the problem, we argue as follows. If $\mathbf{u}_{0}\in\mathcal{S}$
we may not apply the transformation (\ref{eq:transformation-Tminus})
to an entire open neighborhood of $\mathbf{u}_{0}$ in $\mathcal{C}$,
because that neighborhood would necessarily contain points of $\mathcal{C}^{-}$,
where (\ref{eq:transformation-Tminus}) is not defined. Then, if $\mathbf{u}_{0}$
is a grazing, we consider a small open neighborhood $\mathcal{U}\subset\mathcal{S}$
of $\mathbf{u}_{0}$ in $\mathcal{S}$ containing only grazing states
and satisfying $\mathcal{T}^{-}(\mathcal{U})\cap\mathcal{S}=\emptyset$.
Since $R\mathbf{u}=\mathbf{u}$ if $\mathbf{u}\in\mathcal{S},$ we
deduce that the restriction of $\mathcal{T}^{-}$ on $\mathcal{U}$
is as smooth as $\varphi$. If $\mathbf{u}_{0}$ is the beginning
of a sticky event, then we can not consider an open neighborhood of
$\mathbf{u}_{0}$ in $\mathcal{S}$ because it would contain states
having $\tau^{-}=0$, and for those the transformation (\ref{eq:transformation-Tminus})
is not defined. Then we consider open intervals $\tilde{\mathcal{U}}$
along the straight line defined by $\dot{x}=0$, $F=0$. If $\tilde{\mathcal{U}}$
is small enough, then it does not contain states for which $\dot{F}\ge0$
and moreover $\mathcal{T}^{-}(\tilde{\mathcal{U}})\cap\mathcal{S}=\emptyset$.
Then the restriction of $\mathcal{T}^{-}$ on $\tilde{\mathcal{U}}$
is as smooth as $\varphi$.

The union of all the above sets $\mathcal{U}$, $\tilde{\mathcal{U}}$
is \[
\bar{\mathcal{S}}=\left\{ \mathbf{u}\in\mathcal{S}\left|\mathcal{T}^{-}(\mathbf{u})\,\mathrm{is\, defined\, and\,}\mathcal{T}^{-}(\mathbf{u})\notin\mathcal{S}\right.\right\} .\]
 The set $\bar{\mathcal{S}}$ is contained in a two-dimensional plane
and has zero measure in $\mathcal{C}^{+}$, which is three-dimensional,
and the previous discussion proves that $\mathcal{T}^{-}$ is as smooth
as $\varphi$ on $\bar{\mathcal{S}}$. Then the image of $\bar{\mathcal{S}}$
in $\mathcal{C}^{+}$ through $\mathcal{T}^{-}$, namely $\mathcal{T}^{-}(\bar{\mathcal{S}})$,
is also a set of zero measure in $\mathcal{C}^{+}$, see e.g. \cite[Chapter 2, Proposition 1.6]{Golubitsky-Guillemin}.
By analogy, for $n\ge2$, we define \[
\bar{\mathcal{S}}^{n}=\left\{ \mathbf{u}\in\mathcal{T}^{-}(\bar{\mathcal{S}}^{n-1})\left|\mathcal{T}^{-}(\mathbf{u})\notin\mathcal{S}\right.\right\} \]
 and we set $\bar{\mathcal{S}}^{1}=\bar{\mathcal{S}}$, $\bar{\mathcal{S}}^{0}=\mathcal{S}$.
As before, $\mathcal{T}^{-}$ is smooth on $\bar{\mathcal{S}}^{n}$
and then, by induction, $\bar{\mathcal{S}}^{n+1}\subset\mathcal{T}^{-}(\bar{\mathcal{S}}^{n})$
has zero measure in $\mathcal{C}^{+}$ for every $n$. By the sigma-additivity
of the Lebesgue measure, it follows that \[
\Sigma=\cup_{n=0}^{\infty}\bar{\mathcal{S}}^{n}\]
 has zero measure, too.

The bouncing ball starting from an initial condition $\mathbf{u}\in\mathbb{R}^{4}$,
with $\mathbf{u}\cdot\mathbf{e}_{1}>0$, will experience a grazing
or a sticky event if its first contact state $\varphi(\tau(\mathbf{u}),\mathbf{u})$
satisfies \[
\varphi(\tau(\mathbf{u}),\mathbf{u})\in\Sigma.\]
 So, the set $\mathcal{A}$ of anomalous initial conditions that reach
$\mathcal{S}$ after a finite number of contacts is contained in \[
\cup_{t\le0}\varphi(t,\Sigma)=\varphi\left((-\infty,0]\times\Sigma\right)\]
 which has zero measure in $\mathbb{R}^{4}$, by the smoothness of
$\varphi$ with respect to all its variables. 

We now prove that $\mathcal{A}$ is nowhere dense in $\mathbb{R}^{4}$.
Because $\mathcal{A}$ has zero measure, no sphere centered in a point
belonging to $\mathcal{A}$ is a subset of $\mathcal{A}$. Then we
need to show that around each non-anomalous points there is a neighborhood
that does not intersect $\mathcal{A}$. 

Let us assume that $\mathbf{u}\notin\mathcal{A}$ . Let us call $\mathbf{u}(t)$
the state of the system at the time $t$, starting from the initial
condition $\mathbf{u}$. From the results of section \ref{sec:Sticky-Events}
it follows that after $N$ contacts it is $E(\mathbf{u}(t_{N}))<E_{min}$,
where $E_{min}$ is the minimal energy required to have an anomalous
contact (see Lemma (\ref{lem:Min-sticky-energy})). Since the mechanical
energy is a continous function of the state of the system, there is
a neighborhood $\mathcal{V}$ of $\mathbf{u}(t_{N})$ such that $\mathcal{V}\cap\mathcal{S}=\emptyset$
and $E(\mathbf{v})<E_{min}$ if $\mathbf{v}\in\mathcal{V}$. Furthermore,
$\mathcal{V}$ may be taken small enough to ensure that $\mathcal{T}^{-n}(\mathcal{V})\cap\mathcal{S}=\emptyset$
for $n=1,\cdots,N$. Finally, we observe that any neighborhood $\mathcal{U}$
of $\mathbf{u}$ such that $\varphi(\tau(\mathcal{U}),\mathcal{U})\subset\mathcal{T}^{-N}(\mathcal{V})$
does not intersect $\mathcal{A}$.

\section{\label{sec:Numerical-Results}Numerical Results}

\subsection{On the Nature of the Restitution Coefficient}

The dynamics of the mechanical system described in section (\ref{sec:model-description})
may be easily simulated on a digital computer, finding the instants
of impact of the lower point mass with the floor with an accuracy
as high as the machine precision.

If the mechanical system is left free to fall with its internal degree
of freedom not excited (that is with initial conditions $\xi(0)=\dot{\xi}(0)=0$,
$\psi(0)>0,$ and arbitrary $\dot{\psi}(0)$) the resulting dynamics
may, initially, approximate that of the rigid ball model with constant
restitution coefficient (eq. (\ref{eq:standard-model})), provided
that the spring is weakly dissipative and sufficiently rigid, and
that the energy of the initial condition is sufficiently high. An
example is illustrated in figure%
\begin{figure}
\begin{centering}\includegraphics[width=0.9\columnwidth]{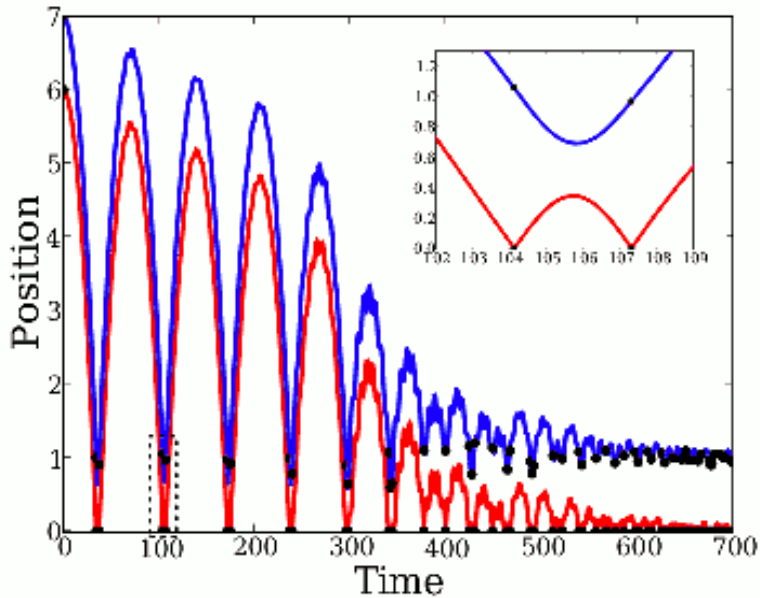} \par\end{centering}

\caption{\label{fig:rigid-fall}Numerical simulation with initial conditions
$\xi(0)=\dot{\xi}(0)=\dot{\psi}(0)=0$, $\psi(0)=6$, and $\gamma=\mu=0.01$.
The upper curve is $y(t)$, the lower curve is $x(t)$. The solid
dots show the instants when there is an impact. The inset magnifies
the region within the dashed rectangle.}
\end{figure}
 (\ref{fig:rigid-fall}), where the mechanical system is left free
to fall from an height of six non-dimensional units.

This fact is explained by observing that, upon an impact, the vibrational
and translational velocities exchange with each other (eq. (\ref{eq:collision-rule})),
so the center of mass of the ball has zero velocity just after the
impact. In this situation we recognize three characteristic times\begin{equation}
T_{\psi}=\sqrt{\frac{1}{2\gamma}},\qquad T_{\xi}=\frac{\pi}{\sqrt{1-\mu^{2}}},\qquad T_{d}=\frac{1}{\mu}.\label{eq:char-times}\end{equation}
 $T_{\psi}$ is the free-fall time of the center of mass, $T_{\xi}$
is half the proper period of the spring, and $T_{d}$ is the e-folding
time associated with energy dissipation from the spring. If $T_{\psi}$
is much larger than $T_{\xi}$ then the center of mass does not have
enough time to gain any appreciable speed before the expanding spring
causes a second impact, which exchanges vibrational and translational
velocities again. After this second impact the mechanical system is
again propelled upward, with a very small vibrational energy left
in the spring. At this point, if the kinetic energy of the center
of mass is high enough to cause a time of flight to be of the order
of $T_{d}$ or larger, then any residual vibrational energy will be
dissipated to negligible levels during the flight, and the mechanical
system will present itself to a third impact with the spring essential
at rest, causing the double-impact dynamics described here to repeat
itself.

In the limit $\gamma\rightarrow0$ the residual vibrational energy
tends to zero, and all dissipation happens during the double impact.
Then it is appropriate to introduce the restitution coefficient \begin{equation}
r=\dot{\Psi}_{n}/\dot{\Psi}_{n+2}\label{eq:proper-rest-coeff}\end{equation}
 where $\dot{\Psi}_{n}$ and $\dot{\Psi}_{n+2}$ are, respectively,
the velocity of the center of mass before and after one of these double-impact
events. It can be shown (see \cite{Pap-Pass} for details) that it
is \begin{equation}
r=e^{-\mu T_{\xi}}.\label{eq:rest-coefficient-function-of-mu}\end{equation}

If we take our mechanical system as model of a bouncing ball, we are
lead to say that during a double-impact event the ball is in contact
with the floor: it takes this finite amount of time for the center
of mass of the ball to exchange momentum with the floor and revert
upward its velocity. Only flights longer than $T_{\xi}$ are to be
taken as macroscopic flights. In the regime discussed here, our model
may be seen as a version of the rigid ball model of eq. (\ref{eq:standard-model})
in which the contacts have a finite, rather than infinitesimal, duration.

When, after a number of double-impact events, the time of flight becomes
shorter than the characteristic dissipation time $T_{d}$, subsequent
impacts will be able to store a significant amount of energy into
the vibrational mode, subtracting it from the energy of the center
of mass. This can be seen clearly in figure (\ref{fig:rigid-fall})
where there is a sudden drop of the maximum height reached by the
mechanical system between the fourth and the sixth macroscopic flight,
and, at the same time, the trajectories of the two point masses become
rather complicated and far from parabolic.

It is interesting a comparison between experiments in which a spherical
bead is free to bounce repeatedly upon a rigid surface until it comes
to rest, and our numerical simulations. Because the velocities of
the center of mass of a bead are difficult to measure, it is customary
in this kind of experiment to measure the time of flight, which is
more accessible to the observer. In the experiments reported in \cite{Falcon98}
the coefficient of restitution is expressed as \begin{equation}
r_{n}=\frac{f_{n+1}}{f_{n}}\label{eq:Falcon-rest-coeff}\end{equation}
 where $f_{n}$ is the time of flight of the bead, (which does not
include the time of contact of the bead with the rigid surface). If
the rigid ball model of eq. (\ref{eq:standard-model}) is valid, then
(\ref{eq:Falcon-rest-coeff}) is equivalent to the standard definition
of restitution coefficient as ratio of velocities before and after
the impact. If the bead is not rigid, then (\ref{eq:Falcon-rest-coeff})
must be taken as an alternate definition of restitution coefficient.
In our numerical simulations, taking into account the interpretation
of double-impacts as contact time with the floor, it is natural to
define \begin{equation}
f_{n}=\tau_{2n}.\label{eq:my-time-of-flight}\end{equation}
 To simulate the very rigid tungsten carbide bead used in \cite{Falcon98}
we set $\gamma=10^{-5}$, while $\mu=0.01$ leads, according to the
approximation (\ref{eq:rest-coefficient-function-of-mu}), to the
restitution coefficient $r=0.969\cdots$, which is about the same
as what is measured in \cite{Falcon98} for relatively high-velocity
impacts. The simulated mechanical system, as the experimental bead,
is left free-falling from a height equal to one quarter of its length.
The restitution coefficient computed from the simulation, according
to the definition (\ref{eq:Falcon-rest-coeff}), is shown in figure
\begin{figure}
\begin{centering}\includegraphics[width=0.9\columnwidth]{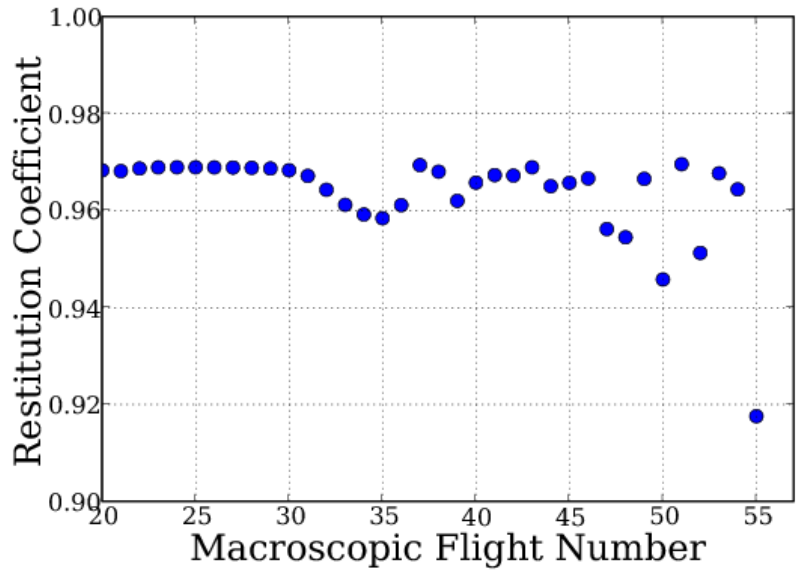} \par\end{centering}

\caption{\label{fig:rest-coeff}Restitution coefficient computed using the
definition (\ref{eq:Falcon-rest-coeff}) from a simulation with $\gamma=10^{-5}$,
$\mu=10^{-2}$, and initial conditions $\xi(0)=\dot{\xi}(0)=\dot{\psi}(0)=0,$
$\psi(0)=0.25$.}
\end{figure}
(\ref{fig:rest-coeff}). Initially the value of $r$ is constant,
and it is in excellent agreement with (\ref{eq:rest-coefficient-function-of-mu}).
Then $r$ declines, and finally it begins to fluctuate without a clear
pattern. At even later times (not pictured in figure) the double-impact
dynamics is completely disrupted, as in the case of figure (\ref{fig:rigid-fall}),
and the definition (\ref{eq:my-time-of-flight}) ceases to be meaningful.
This kind of behavior is qualitatively the same as that of the bead
observed in \cite{Falcon98} (see their figure (7)). In their case,
after fluctuations in the value of $r$ of the same order of magnitude
as those of figure (\ref{fig:rest-coeff}), the bead is observed to
vibrate while maintaining contact with the floor. We stress that the
behavior of figure (\ref{fig:rest-coeff}) remains qualitatively the
same for any choice of parameters corresponding to slightly damped,
very rigid springs.

A different way to look at the same dynamics is pictured in figure
\begin{figure}
\begin{centering}\includegraphics[width=0.9\columnwidth]{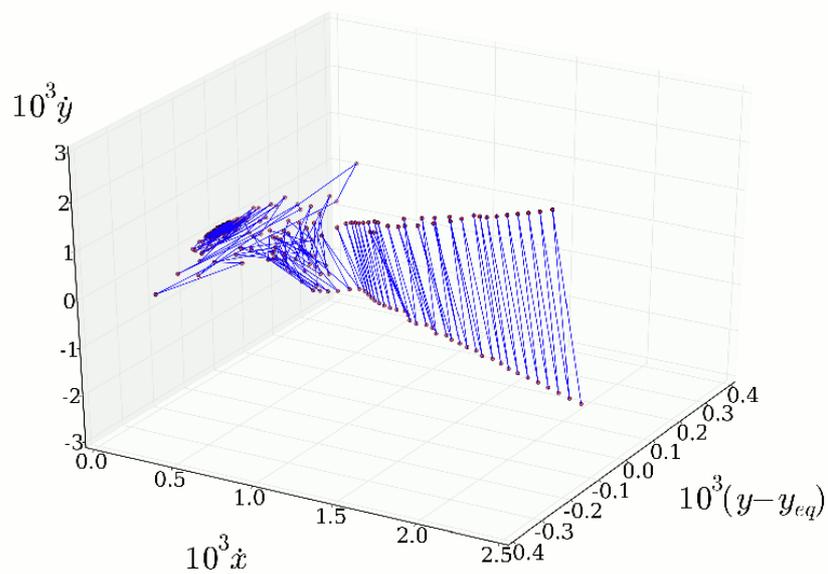} \par\end{centering}

\caption{\label{fig:very_rigid_poincare}The dots represent the state of the
system immediately after a contact with the floor (that is, $x=0$)
for the simulation of figure (\ref{fig:rest-coeff}). The equilibrium
position of the upper point mass is $y_{eq}=1-2\gamma$. The thin
lines have the sole purpose of making evident the temporal sequence
of states.}
\end{figure}
 (\ref{fig:very_rigid_poincare}) where we show the sequence of states
belonging to the set $\mathcal{C}$ of contact configurations (\ref{eq:C})
for the same simulation that generated figure (\ref{fig:rest-coeff}).
Initially, the dynamics is clearly dominated by double-impacts, whose
signature is the alternation of negative and positive values of $\dot{y}$,
which gives a characteristic zig-zag look at the early part of the
sequence. At later times the sequence of states becomes very disordered,
and lacks any easily recognizable pattern, except a tendency to move
towards the equilibrium point. Figure%
\begin{figure}
\begin{centering}\includegraphics[width=0.9\columnwidth]{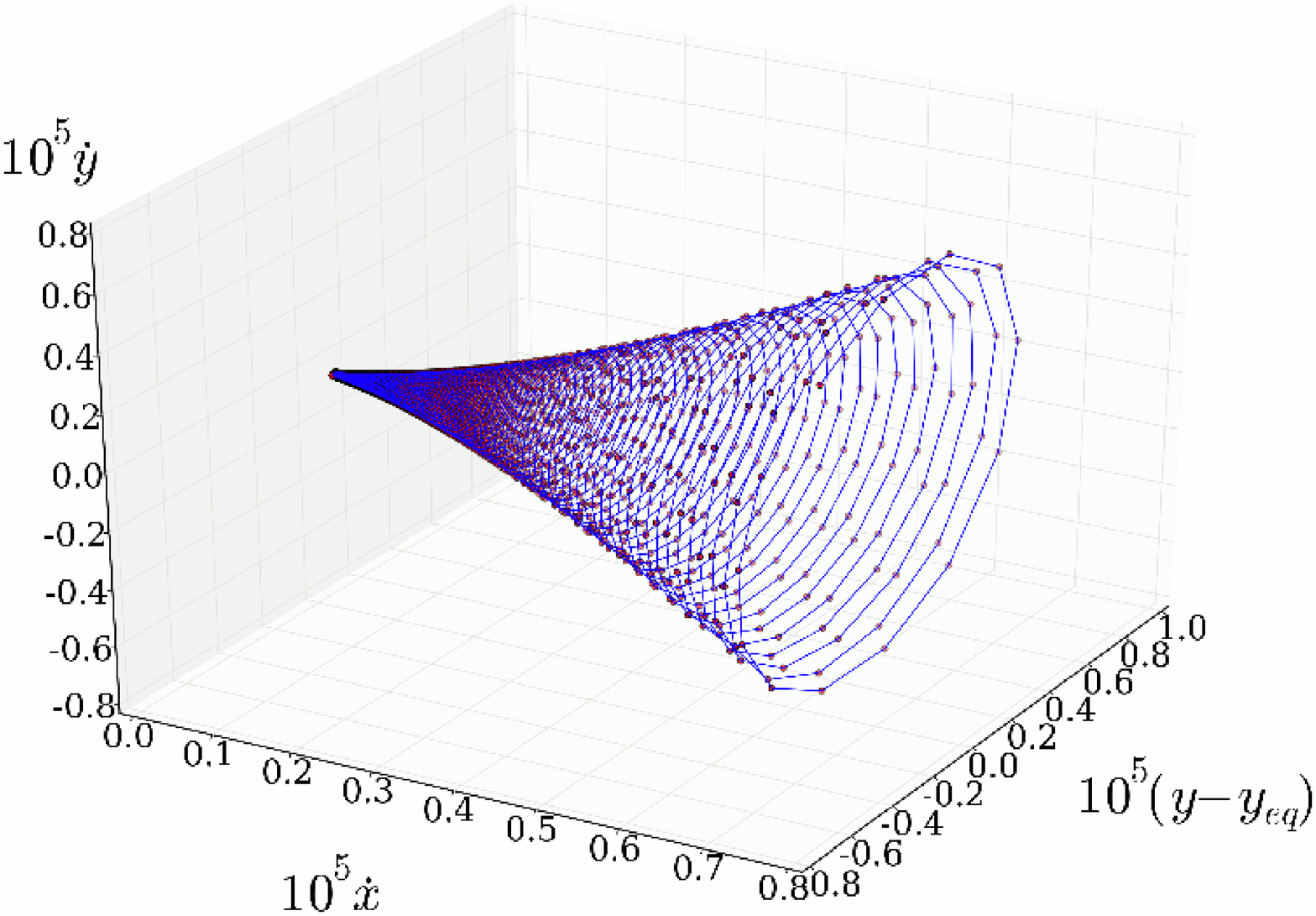} \par\end{centering}

\caption{\label{fig:very_rigid_poincare_magni}Magnification of the figure
(\ref{fig:very_rigid_poincare}), close to $(0,0,0)$.}
\end{figure}
 (\ref{fig:very_rigid_poincare_magni}) is a magnification of figure
(\ref{fig:very_rigid_poincare}) close to the equilibrium point. At
these even later stages, the dynamics is well approximated by the
map (\ref{eq:mappa}). The length of the spring and the velocity of
the upper point mass are both subject to damped oscillations around
the equilibrium position, and this gives a spiraling appearance to
the sequence of states.

The reader may have noticed that the asymptotics of Theorem (\ref{thm:main_thm})
are independent of the particular value of the damping parameter $\mu$.
In particular, the results of the theorem apply equally to under-damped
and over-damped oscillations. In figures %
\begin{figure}
\begin{centering}\includegraphics[width=0.9\columnwidth]{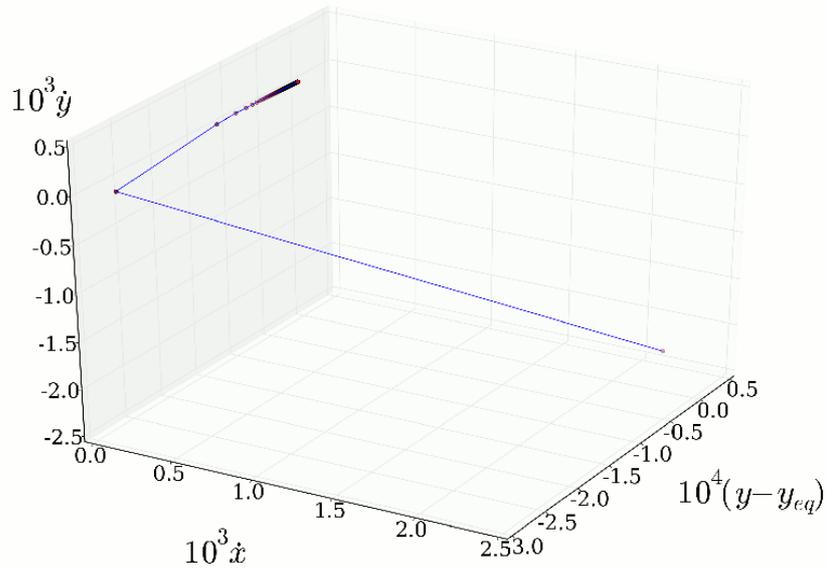} \par\end{centering}

\caption{\label{fig:very_rigid_poincare_overdamped}Same as figure (\ref{fig:very_rigid_poincare}),
but with $\mu=2$.}
\end{figure}
 (\ref{fig:very_rigid_poincare_overdamped}) and %
\begin{figure}
\begin{centering}\includegraphics[width=0.9\columnwidth]{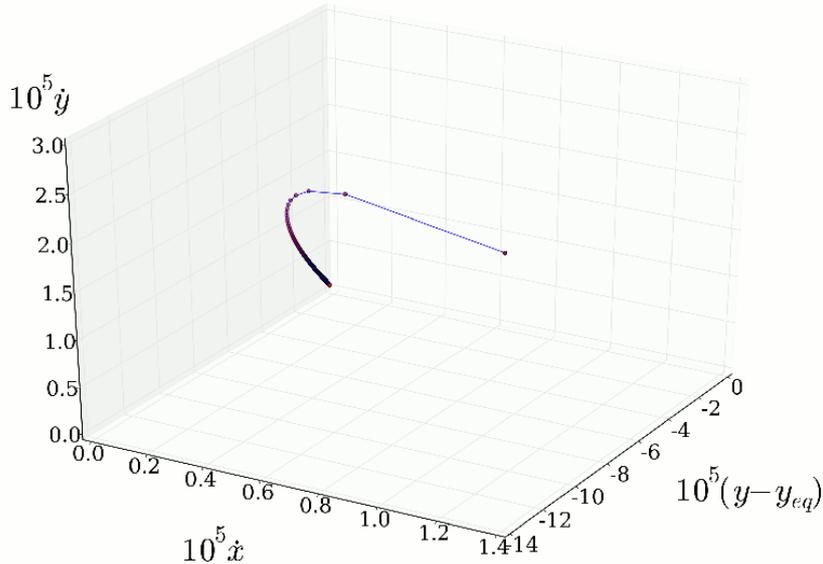} \par\end{centering}

\caption{\label{fig:very_rigid_poincare_overdamped_magni} Magnification of
figure (\ref{fig:very_rigid_poincare_overdamped}), close to $(0,0,0)$.}
\end{figure}
 (\ref{fig:very_rigid_poincare_overdamped_magni}) we show a simulation
that uses $\gamma=10^{-5}$, $\mu=2$ and the same initial conditions
as the simulation of figure (\ref{fig:very_rigid_poincare}). The
length of the overdamped spring shrinks for the first few impacts,
then slowly expands, approaching the equilibrium value monotonically
from below. The system enters immediately the asymptotic regime, and
the is no disordered transient. From a macroscopic point of view,
the idealized ball may be taken as performing a single, totally inelastic
impact with the floor.

\subsection{Sticky Solutions}

We have performed several numerical simulations using the following
initial conditions\begin{equation}
\left\{ \begin{array}{l}
X_{0}=0\\
\dot{X}_{0}=\epsilon\\
Y_{0}=1+2\gamma-2\mu\dot{Y}_{0}\\
\dot{Y}_{0}=-0.1\end{array}\right..\label{eq:sticky-ic}\end{equation}
 For $\epsilon=0$ they obey the conditions (\ref{eq:sticky-cond-Y}),
leading to a sticky event which lasts for $t_{c}=5.23\cdots$ non-dimensional
time units. In figure %
\begin{figure}[p]

\begin{centering}\includegraphics[width=0.8\columnwidth]{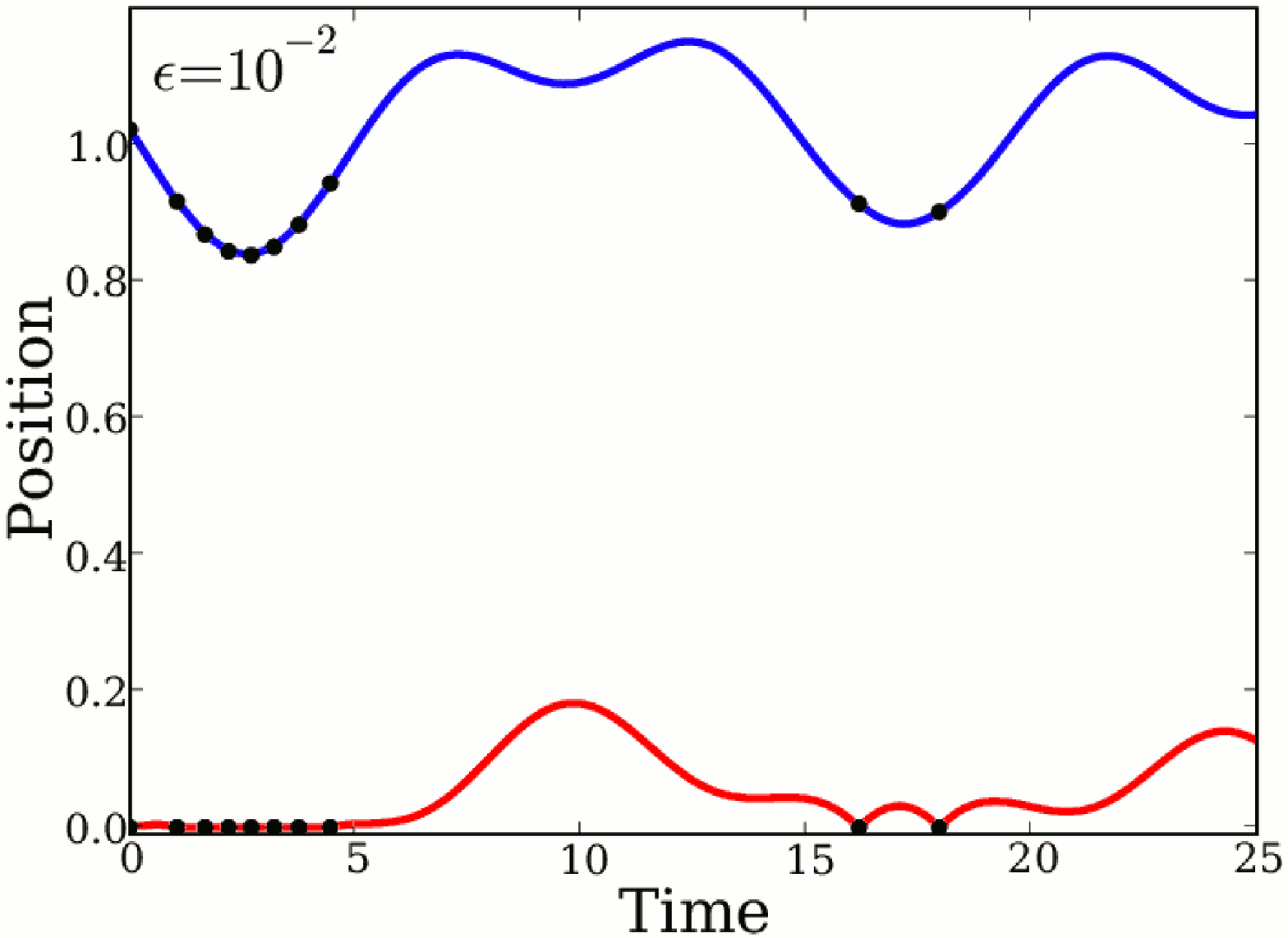} \par\end{centering}

\begin{centering}\includegraphics[width=0.8\columnwidth]{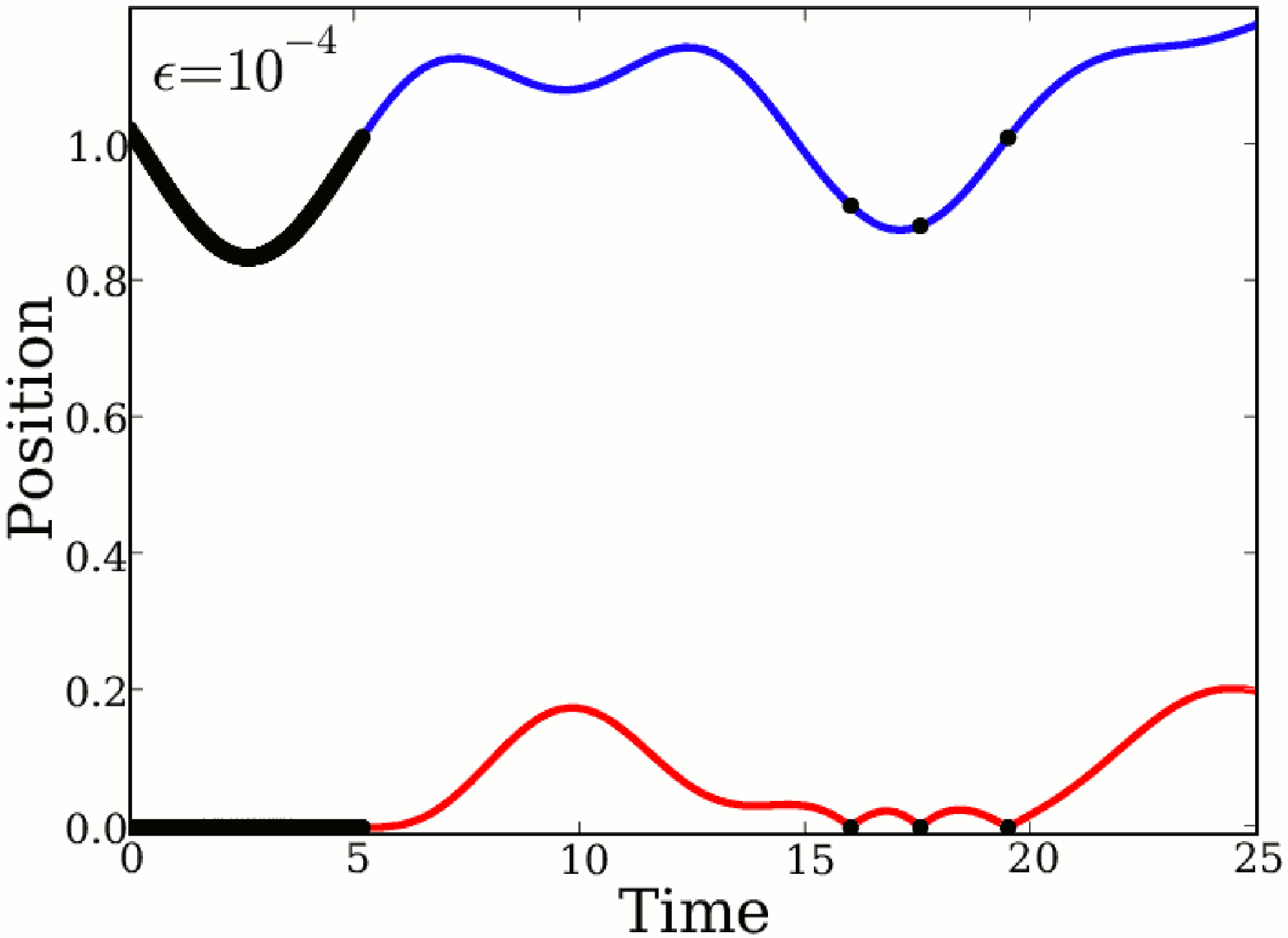} \par\end{centering}

\begin{centering}\includegraphics[width=0.8\columnwidth]{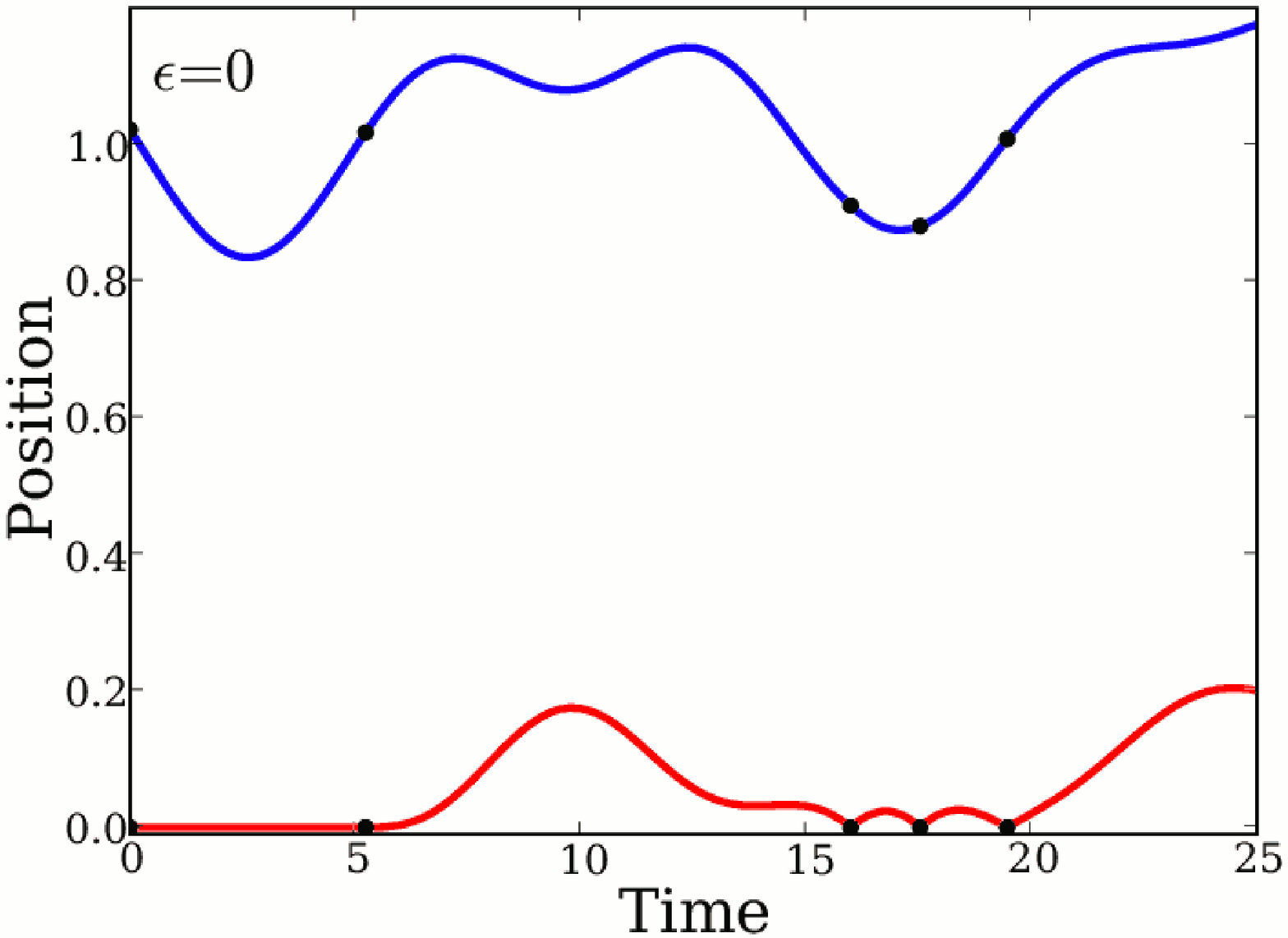} \par\end{centering}

\caption{\label{fig:sticky-traj}Simulations using the initial conditions
(\ref{eq:sticky-ic}) for three different values of $\epsilon$. Solid
dots mark the times of impact. In the panel $\epsilon=0$, in the
interval between the first two solid dots, the lower point mass is
in contact with the floor.}
\end{figure}
 (\ref{fig:sticky-traj}) some sample trajectories are shown. It appears
that, as $\epsilon$ decreases, the solutions tend to the solution
with $\epsilon=0$. To quantify the convergence, we have computed
the norm\begin{equation}
\left\Vert y_{\epsilon}-y_{s}\right\Vert =\sqrt{\frac{1}{t_{c}}\int_{0}^{t_{c}}(y_{\epsilon}(t)-y_{s}(t))^{2}dt}\label{eq:L2-norm}\end{equation}
 where $y_{\epsilon}$ and $y_{s}$ are, respectively, the position
of the upper point mass with $\epsilon\neq0$, and with $\epsilon=0$.
The results are shown in figure %
\begin{figure}
\begin{centering}\includegraphics[width=0.9\columnwidth]{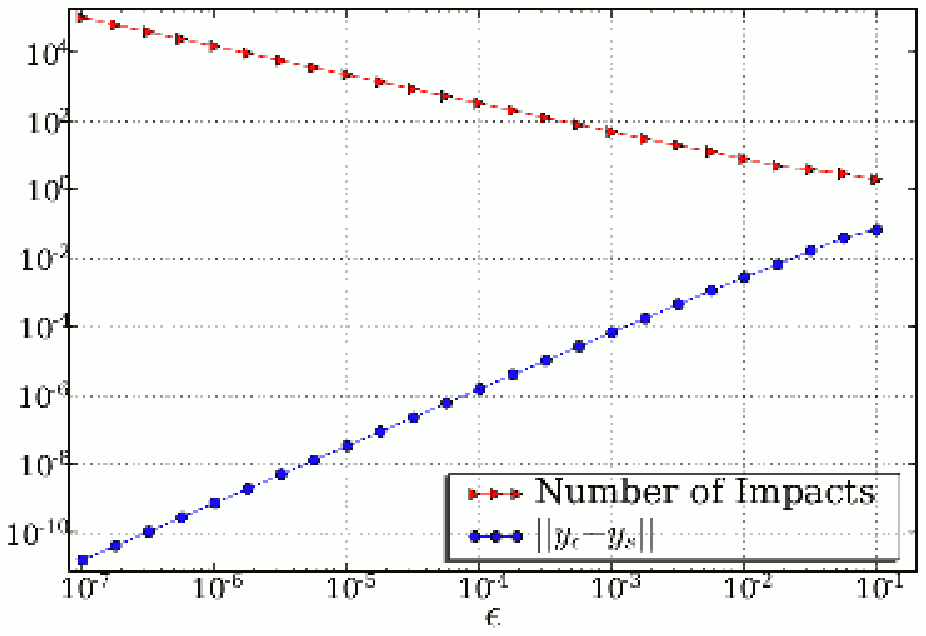} \par\end{centering}

\caption{\label{fig:sticky-convergence}Number of impacts performed by the
lower point mass in the interval $(0,t_{c})$ and root mean squared
difference (\ref{eq:L2-norm}) as function of $\epsilon$.}
\end{figure}
 (\ref{fig:sticky-convergence}). The root-mean-squared difference
(\ref{eq:L2-norm}) falls to zero faster than linearly with $\epsilon$.

The numerical evidence supports the idea that the dynamics during
a sticky event may be approximated up to any degree of accuracy by
a non-sticky dynamics.

\section{\label{sec:Discussion-and-Conclusions}Discussion and Conclusions}

Our simple model shows that taking into account explicitly the deformability
of a bouncing body avoids the pathology of the inelastic collapse.
Of course this comes at a price: our description of the bouncing ball,
in spite of its simplicity, is considerably more complicated than
a restitution-coefficient based model. This is affordable for one,
or maybe a few, particles, but it is unbearable for systems with a
very large number of bodies, even resorting to numerical simulation.

One may be tempted to overcome this difficulty by returning to a description
based upon a restitution coefficient of the form\begin{equation}
r(u)=\left\{ \begin{array}{lr}
\exp\left(-\frac{\mu\pi}{\sqrt{1-\mu^{2}}}\right), & u>u_{c}\\
1-\frac{\mu}{3\gamma}u, & u\le u_{c}\end{array}\right.\label{eq:our-r}\end{equation}
where $u_{c}=3\gamma\mu^{-1}\left(1-\exp(-\mu\pi/\sqrt{1-\mu^{2}})\right)$.
This is obtained by stitching the expression (\ref{eq:rest-coefficient-function-of-mu}),
which is valid for times of flight larger than the characteristic
damping time of internal vibrations, and the coefficient in expression
(\ref{eq:restituzione}), which is the map that approximates the asymptotic
vibrations of a ball close to equilibrium.

Although (\ref{eq:our-r}) is very accurate for large and for small
values of $u$, a glance at the figures of section \ref{sec:Numerical-Results}
shows that (for underdamped springs) there is an intermediate range
of impact speeds where (\ref{eq:our-r}) simply fails. When the internal
vibrational mode is excited, but the times of flight are not yet infinitesimal,
the dynamics is too irregular to be parameterized with a velocity-dependent
restitution coefficient.

We may hint that a satisfactory parameterization will need some form
of bookkeeping of the energy stored in the internal vibrations, and
some rule to determine the exchanges between that and the energy of
the center of mass. At this stage, we are unable to suggest anything
more specific than these vague considerations. In particular, it is
not at all clear if accounting for internal vibrations will help the
ongoing search for some sort of hydrodynamic limit for systems with
a large number of dissipative particles.

We remark that the behavior of the bouncing ball, close to equilibrium,
does not depend crucially on the linearity of the spring. A nonlinear
variant of the model described in section (\ref{sec:model-description})
is embodied by the equations\begin{equation}
\left\{ \begin{array}{c}
\ddot{\psi}=-\gamma\\
\ddot{\xi}=-\rho\xi\left|\xi\right|^{a}-2\mu\dot{\xi}\left|\xi\right|^{b}\end{array}\right.\label{eq:eq-motion-nonlin-psi-xi}\end{equation}
 with $\rho,\mu>0$, $a,b\ge0$. The impact times $t_{n}$ are determined
by condition (\ref{eq:collision-condition-psi-xi}), where the collision
rule (\ref{eq:collision-rule}) must be applied. The constraint $\psi(t)>\xi(t)$
holds at all times. We recover the linear model for $a=b=0$. It has
been argued, on the basis of Hertz's contact law, that the correct
choice for modeling homogeneous spheres is $a=b=1/2$, and that the
restitution coefficient (\ref{eq:onefifth-rest-coeff}) may be justified,
for vanishing speeds of impact, by computing the energy dissipated
by (\ref{eq:eq-motion-nonlin-psi-xi}) in a single compression-expansion
cycle (see \cite{Brilliantovbook} chap. 3 and references therein).
Although some of the claims in the literature appear to be questionable
in their generality, because the vector field defining (\ref{eq:eq-motion-nonlin-psi-xi})
is non-Lipschitz for $\xi=0$ and $0<b<1$, the problem is well-posed
close to the position of static equilibrium, which is \begin{equation}
\psi=\dot{\psi}=\dot{\xi}=0,\quad\xi=-\left(\frac{\gamma}{\rho}\right)^{\frac{1}{a+1}}.\label{eq:nonlin-equilibrium-xi}\end{equation}
 That corresponds to the minimum of the energy\begin{equation}
E=\frac{\dot{\psi}^{2}}{2}+\frac{\dot{\xi}^{2}}{2}+\frac{\rho}{a+2}\left|\xi\right|^{a+2}+\gamma\psi\label{eq:nonlin-energy}\end{equation}
 which is a non-increasing function of time:\[
\frac{dE}{dt}=-\mu\dot{\xi}^{2}\left|\xi\right|^{b}.\]
 If we take a contact configuration $(0,\dot{X}_{0},Y_{0},\dot{Y}_{0})$
close enough to the static equilibrium (\ref{eq:nonlin-equilibrium-xi})
the system remains close to the minimum of the energy at all later
times. As a consequence, we are guaranteed that $\ddot{X}_{n}$ and
$\dddot{X}_{n}$ are in a neighborhood of $-2\gamma$ and $2\mu\gamma(\gamma/\rho)^{b/(a+1)}$,
respectively. Then we follow the steps (\ref{eq:x-exact-expansion})
through (\ref{eq:xdot_map}), using (\ref{eq:tau3}) as the time of
flight, and we find that the asymptotic dynamics follows the map (\ref{eq:mappa})
with $\alpha_{n}\rightarrow\mu(3\gamma)^{-1}(\gamma/\rho)^{b/(a+1)}$
for $n\rightarrow\infty$. Then we conclude that a sequence of repeated
impacts, because of the deformability of the body, follows the same
asymptotic law as the linear case. As a consequence, the restitution
coefficient (\ref{eq:onefifth-rest-coeff}) may be accurate only in
the case of well-separated impacts, that is, when the time of flight
between consecutive impacts is large enough to allow for dissipation
of internal vibrations.

\section{Appendix}

\begin{lem}
\label{lem:Giorgio's}Given the map \[
x_{n+1}=x_{n}-\alpha_{n}x_{n}^{2}+o(x_{n}^{2})\]
 with $0<x_{0}\ll1$ and $\lim_{n\rightarrow\infty}\alpha_{n}=\alpha>0$,
then\[
\lim_{n\to\infty}x_{n}=0\quad and\quad\lim_{n\to\infty}nx_{n}=\frac{1}{\alpha}.\]

\end{lem}
\begin{proof}
Since $(\alpha_{n}-\alpha)x_{n}^{2}=o(x_{n}^{2})$, we may write\begin{equation}
x_{n+1}=x_{n}-\alpha x_{n}^{2}+o(x_{n}^{2}).\label{eq:x-x2-map}\end{equation}
 For small $x_{0}$, the sequence generated by (\ref{eq:x-x2-map})
is positive, decreasing and $\lim_{n\to\infty}x_{n}=0$. For a given
$\epsilon$, let us take $N$ such that for any $n>N$\begin{equation}
1-(a+\epsilon)x_{n}\leq\frac{x_{n+1}}{x_{n}}\leq1-(a-\epsilon)x_{n}\label{eq:ineq1}\end{equation}
 We may assume $x_{N}<1/(\alpha-\epsilon)$ , then $x_{N+k}<1/(\alpha-\epsilon)$;
we prove that $x_{N+k}\le1/k(\alpha-\epsilon)$ implies $x_{N+k+1}\le1/(k+1)(\alpha-\epsilon)$
and apply induction. Multiplying the right inequality in \eqref{eq:ineq1}
by $1+(a-\epsilon)x_{n}$, we have\[
\frac{x_{n+1}}{x_{n}}(1+(a-\epsilon)x_{n})\leq1\]
 Then \[
x_{N+k+1}\leq\frac{x_{N+k}}{1+(a-\epsilon)x_{N+k}}\]
 where $x/1+(a-\epsilon)x$ is an increasing function of $x$ and
$x_{N+k}\le1/k(\alpha-\epsilon)$, then

\[
x_{N+k+1}\leq\frac{\frac{1}{k(\alpha-\epsilon)}}{1+(a-\epsilon)\frac{1}{k(\alpha-\epsilon)}}=\frac{1}{(k+1)(\alpha-\epsilon)}.\]
 By induction

\[
x_{n}\le\frac{1}{(n-N)(\alpha-\epsilon)}\quad\forall n>N\]
 then

\[
\limsup_{n\rightarrow\infty}nx_{n}\le\frac{1}{\alpha-\epsilon}\qquad\forall\varepsilon>0\]
 i.e.

\[
\limsup_{n\to\infty}nx_{n}\le\frac{1}{\alpha}.\]
 By a similar argument we get\[
\liminf_{n\to\infty}nx_{n}\ge\frac{1}{\alpha}\]
 which proves the proposition. 
\end{proof}
\begin{rem}
\label{rem:x-xbeta-map} It follows that for a map $x_{n+1}=x_{n}-\alpha x_{n}^{\beta}+o(x_{n}^{\beta})$
, with $\beta>1$, \begin{equation}
\lim_{n\to\infty}x_{n}=0\quad{\normalcolor \mathrm{and}}\quad\lim_{n\to\infty}n^{\frac{1}{\beta-1}}x_{n}>0.\label{eq:remark-xbeta}\end{equation}
 This can be seen by using the transformation $z_{n}=x_{n}^{\beta-1}$.
We have \[
\frac{z_{n+1}}{z_{n}}=\left(1-\alpha x_{n}^{\beta-1}+o(x_{n}^{\beta-1})\right)^{\beta-1}=1-\alpha(\beta-1)x_{n}^{\beta-1}+o(x_{n}^{\beta-1})\]
 We then apply Lemma \ref{lem:Giorgio's} to \[
z_{n+1}=z_{n}-\alpha(\beta-1)z_{n}^{2}+o(z_{n}^{2})\]
 which gives (\ref{eq:remark-xbeta}). 
\end{rem}
\begin{lem}
\label{lem:Giorgio's-2}If $x_{n}>0$ and $x_{n+1}=x_{n}+O(x_{n}^{2})$
then $\sum_{n}x_{n}=+\infty$. 
\end{lem}
\begin{proof}
We have $x_{n+1}\ge x_{n}-cx_{n}^{2}$ for a suitable $c>0$. Hence,
if we define \[
\left\{ \begin{array}{ccc}
z_{n+1} & = & z_{n}-cz_{n}^{2}\\
z_{0} & = & x_{0}\end{array}\right.\]
 we have $x_{n}\ge z_{n}$ for all $n\ge0$. Applying Lemma (\ref{lem:Giorgio's})
to this map, the thesis follows. 
\end{proof}
\begin{lem}
\label{lem:alpha-map}If there exists a sequence $\left\{ \alpha_{n}\right\} _{n\in\mathbb{N}}$
with $\alpha_{n}\in(0,1)$ that satisfies (\ref{eq:alpha-map}), then
$\lim_{n\rightarrow\infty}\alpha_{n}=0$. 
\end{lem}
\begin{proof}
Let us define the functions $f$ and $g$ as\begin{equation}
f(x)=6x-4x^{2}+x^{3},\label{eq:f}\end{equation}
 \begin{equation}
g(x)=\frac{6x-8x^{2}+3x^{3}}{\left(1-x\right)^{3}}.\label{eq:g}\end{equation}
 The map (\ref{eq:alpha-map}) is then written as \begin{equation}
f(\alpha_{n+1})=g(\alpha_{n})+b_{n},\label{eq:fg-map}\end{equation}
 with $b_{n}\to0$. Since $f$ is bounded in $[0,1]$ and $g$ tends
to $\infty$ as $x\to1$, it is easy to see that there exists $\delta>0$
such that $\alpha_{n}\le1-\delta$. We observe, that $f^{\prime}(x)>0$.
With some algebra, we also find, for $x\in[0,1)$, \begin{equation}
f(x)-g(x)\le-\frac{x^{2}}{(1-x)^{3}}.\label{eq:g-minus-f}\end{equation}
 In particular, the map (\ref{eq:alpha-map}) has no fixed point in
$(0,1)$. Let us assume that $\{\alpha_{n}\}$ does not converge to
zero. This means $\alpha_{n}\ge\epsilon$ for infinitely many values
of $n$ and some $\epsilon>0$. Inequality (\ref{eq:g-minus-f}) implies
that $g(x)-f(x)\le-\epsilon^{2}$ for $x\ge\epsilon$. Let us fix
$m$ such that $|b_{n}|\le\epsilon^{2}/2$ for $n\geq m$ and take
any $\alpha_{n}\geq\epsilon$ with $n\ge m$. Then the inequality
$\alpha_{n+1}\le\alpha_{n}$ would give, since $f$ is increasing,
\[
b_{n}=f(\alpha_{n+1})-g(\alpha_{n})\le f(\alpha_{n})-g(\alpha_{n})\le-\epsilon^{2}\]
 which contradicts the choice of $b_{n}$. Therefore $\alpha_{n+1}>\alpha_{n}\geq\epsilon$
and, iterating this argument, the sequence $\alpha$ is eventually
increasing and convergent to a number $0<\alpha<1$ satisfying $f(\alpha)=g(\alpha)$.
Since this contradicts (\ref{eq:g-minus-f}), the proof is complete. 
\end{proof}


\begin{thebibliography}{10}
\bibitem{Cross}R. Cross, \emph{The Coefficient of Restitution for
Collisions of Happy Balls, Unhappy Balls, and Tennis Balls}, Am. J.
Phys., \textbf{68}, 1025-1031 (2000).

\bibitem{GuckHol}J. Guckenheimer, P. Holmes, \emph{Nonlinear Oscillations,
Dynamical Systems and Bifurcations of Vector Fields}, Springer, Berlin/Heidelberg
(1983).

\bibitem{Brilliantovbook}N. V. Brilliantov, T. Pöschel, \emph{Kinetic
Theory of Granular Gases}, Oxford University Press, Oxford, (2004).

\bibitem{McNamara-Young}S. McNamara and W. R. Young, \emph{Inelastic
Collapse and Clumping in a One-Dimensional Granular Medium}, Phys.
Fluids A, \textbf{4}, 496-504, (1992).

\bibitem{SchorghoferZhou}N. Schörghofer and T. Zhou, \emph{Inelastic
Collapse of Rotating Spheres}, Phys. Rev. E, \textbf{54}, 5511-5515,
(1996).

\bibitem{Ramirez}R. Ramírez, T. Pöschel, N. V. Brilliantov, and T.
Schwager, \emph{Coefficient of restitution of colliding viscoelastic
spheres}, Phys. Rev. E, \textbf{60}, 4465-4472, (1999).

\bibitem{Bridges84}F. G. Bridges, A. Hatzes, and D. N. Lin, \emph{Structure,
Stability and Evolution of Saturn's Rings}, Nature, 309, 333-335,
(1984).

\bibitem{Goldman}D. Goldman, M. D. Shattuck, C. Bizon, W. D. McCormick,
J. B. Swift, and H. L. Swinney, \emph{Absence of Inelastic Collapse
in a Realistic Three Ball Model}, Phys. Rev. E, \textbf{57}, 4831-4833,
(1998).

\bibitem{Duran99}J. Duran, Sands, \emph{Powders, and Grains: An Introduction
to the Physics of Granular Materials}, Springer, Berlin/Heidelberg
(1999).

\bibitem{Luding-McNamara}S. Luding, S. McNamara, \emph{How to Handle
the Inelastic Collapse of a Dissipative Hard-Sphere Gas with the TC
Model}, Granular Matter, \textbf{1}, 113-128, (1998).

\bibitem{Zippelius}T. Aspelmeier, A. Zippelius, \emph{Dynamics of
a One-Dimensional Granular Gas with a Stochastic Coefficient of Restitution},
Physica A, \textbf{282}, 450-474, (2000).

\bibitem{Budd-book}M. di Bernardo, C. Budd, A. Champneys, P. Kowalczyk,
\emph{Piecewise-smooth Dynamical Systems: Theory and Applications},
Springer, Berlin/Heidelberg, 2007.

\bibitem{Falcon98}E. Falcon et al., \emph{Behavior of one inelastic
ball bouncing repeatedly off the ground}; Eur. Phys. J. B, \textbf{3},
45-57, (1998).

\bibitem{Golubitsky-Guillemin}M. Golubitsky, V. Guillemin, Stable
Mappings and Their Singularities, Springer, Berlin/Heidelberg, 1973.

\bibitem{Pap-Pass}F. Paparella and G. Passoni, \emph{Absence of inelastic
collapse for a one-dimensional gas of grains with an internal degree
of freedom}; Comp. Math. App. (2007), in press. 
\end{thebibliography}
\end{document}